\documentclass[12pt,a4paper]{article}

\usepackage{jheppub}
\usepackage[utf8]{inputenc}
\usepackage[english]{babel}
\usepackage{color}
\usepackage{amsmath}
\usepackage{verbatim}   
\usepackage{subfigure}  
\usepackage{amsfonts}
\usepackage{amssymb}
\usepackage{mathrsfs}
\usepackage{stmaryrd}
\usepackage{graphicx}
\usepackage{slashed}
\usepackage{amsthm}
\usepackage{graphicx}
\usepackage{dcolumn}
\usepackage{bm}
\usepackage{color}
\usepackage{amsmath}
\usepackage{verbatim}   
\usepackage{subfigure}  
\usepackage{amsfonts}
\usepackage{amssymb}
\usepackage{mathrsfs}
\usepackage{slashed}
\usepackage{amsthm}
\usepackage{cleveref}
\usepackage[]{units} 

\def\d{\text{d}}
\newcommand{\svec}[1]{\, \text{\tiny $\left(\begin{array}{c} #1 \end{array}\right)$}\, }
\newcommand{\mvec}[1]{\, \text{\small $\left(\begin{array}{c} #1 \end{array}\right)$}\, }
\newcommand{\sfrac}[2]{\sqrt{\frac{#1}{#2}}}

\makeatletter   
\renewcommand*\env@matrix[1][*\c@MaxMatrixCols c]{%
  \hskip -\arraycolsep
  \let\@ifnextchar\new@ifnextchar
  \array{#1}}
\makeatother

\newenvironment{psmallmatrix}
  {\left(\begin{smallmatrix}}
  {\end{smallmatrix}\right)}

\begin{document}

\title{Dressed elliptic genus of heterotic compactifications with torsion and general bundles}
\author[a,b]{Dan Isra\"el}
\author[a,b]{Matthieu Sarkis}

\affiliation[a]{Sorbonne Universit\'es, UPMC Univ Paris 06, UMR 7589, LPTHE, F-75005, Paris, France}
\affiliation[b]{CNRS, UMR 7589, LPTHE, F-75005, Paris, France}

\emailAdd{israel@lpthe.jussieu.fr}
\emailAdd{msarkis@lpthe.jussieu.fr}

\null\vskip10pt

\date{\today}
\abstract{We define and compute the dressed elliptic genus of $\mathcal{N}=2$ heterotic compactifications with torsion that are 
principal two-torus bundles over a K3 surface. We consider a large class of gauge bundles compatible with supersymmetry, 
consisting of a stable holomorphic vector bundle over the base together with an Abelian 
bundle over the total space, generalizing the computation previously done by the authors in the absence 
of the latter. Starting from a (0,2) gauged linear sigma-model with torsion we use supersymmetric 
localization to obtain the result. We provide also a mathematical definition of the dressed elliptic genus as a modified 
Euler characteristic and prove that both expressions agree for hypersurfaces in weighted projective spaces. Finally we 
show that it admits a natural decomposition in terms of $\mathcal{N}=4$ superconformal characters, that may be useful to 
investigate moonshine phenomena for this wide class of $\mathcal{N}=2$ vacua, that includes 
$K3\times T^2$ compactifications as special cases.  
}

\maketitle


\newpage

\section{Introduction}

The supersymmetry constraints at order $\alpha'$ leading to heterotic compactifications with $\mathcal{N}=1$ supersymmetry in 
four space-time dimensions have been known since the seminal works of Hull~\cite{Hull:1986kz} and 
Strominger~\cite{Strominger:1986uh}, and are summarized in a set of BPS equations known as the Strominger system. Solutions of 
this system consisting of a Calabi-Yau 3-fold whose spin connection is embedded in the gauge connection, 
or Calabi-Yau 3-folds equipped with more general gauge bundles, have been extensively studied 
in the context of string phenomenology. However, such solutions also come with a collection of moduli which 
are phenomenologically undesirable. 

A standard approach to fix part of the moduli is to turn on fluxes (three-form flux $H$, playing the role of totally 
antisymmetric torsion, in the heterotic case) through the cycles of the internal geometry. Very few heterotic compactifications 
admitting non-vanishing torsion  are known. The main reasons are that 
whenever $H\neq 0$, the geometry is no longer K\"ahler, and that the Bianchi identity is non-linear in the flux. 

A well-studied class of solutions, leading to $\mathcal{N}=1$ or $\mathcal{N}=2$, consists in $T^2$ principal bundles over 
warped $K3$ surfaces. These solutions were first discovered by Dasgupta, Rajesh and Sethi from type IIB orientifolds via 
S-duality \cite{Dasgupta:1999ss}, and their $SU(3)$ structure was obtained by Goldstein and Prokushkin \cite{Goldstein:2002pg}. 
Later, Fu and Yau solved the Bianchi identity~\cite{Fu:2006vj}, which reduces to a partial differential equation for 
a single function, using the Chern connection on the tangent bundle. As argued in~\cite{Becker:2009df}, another choice 
of connection should be used if one wants to avoid  corrections to the BPS equations at order $\alpha'$. 

One of the main technical advantage of heterotic strings is that they allow for a worldsheet description of fluxes, 
compared to type II superstrings for which no usable worldsheet description of the various Ramond-Ramond fluxes is known; 
the generic lack of large-volume limit in the heterotic case makes this approach necessary anyway. A very fruitful 
approach in studying the worldsheet theory is to find a simple theory flowing in the IR to the superconformal 
non-linear sigma-model of interest. Quantities which are invariant under the RG flow, for 
instance quantities depending only on the topology of the  target space, can be  evaluated more easily.

The gauged linear sigma model (GLSM) approach, proposed initially by Witten to describe Calabi-Yau 
compactifications~\cite{1993NuPhB.403..159W}, provides such UV completion of the worldsheet theory in terms 
of a gauge theory, usually Abelian, with $(2,2)$ or $(0,2)$ supersymmetry. Moreover, as was discussed 
in~\cite{Silverstein:1995re,Beasley:2003fx,Bertolini:2014dna}, at least a large class of 
those should be free of destabilization by worldsheet instantons. This construction was then extended to the type of 
non-K\"ahler $\mathcal{N}=2$ compactifications considered here in~\cite{Adams:2006kb}, where 
torsional GLSMs (TGLMs) were introduced.\footnote{Torsional GLSMs describing other types of compactifactions with 
flux are discussed in~\cite{Adams:2009av,2011JHEP...03..045A,Blaszczyk:2011ib,Quigley:2011pv,Quigley:2012gq,Adams:2012sh}.} 
These torsional GLSMs were used to compute the massless spectrum of the underlying flux 
compactification using Landau-Ginzburg methods in~\cite{2011JHEP...03..045A}, and to study their properties under 
T-duality in~\cite{2013JHEP...11..093I}. 

This torsion GLSM approach was further exploited by the present authors to compute their new supersymmetric 
index~\cite{Israel:2015aea} using supersymmetric localization, building upon techniques 
developed in~\cite{Gadde:2013dda,2014LMaPh.104..465B,2015CMaPh.333.1241B} for elliptic genera, and extending 
known results for $K3\times T^2$ solutions to this more general class of 
$\mathcal{N}=2$ compactifications. The computation involved a modified elliptic genus suitable for the non-K\"ahler geometries 
$T^2\hookrightarrow X\rightarrow K3$, consisting in a non-holomorphic (in $\tau$) dressing of the anomalous elliptic 
genus of the base by the lattice of the torus fiber.
We will refer to this supersymmetric index as the \textit{dressed elliptic genus} in the following. A geometrical definition of this index was given, independent of any underlying $2$-dimensional model. 

However, in the aforementioned article we considered only gauge bundles which are pullbacks of 
stable holomorphic bundles over the $K3$ base. It is known~\cite{Becker:2006et} that 
an additional Abelian gauge bundle over the total space of the principal $T^2$ bundle, that would reduce 
to a set of Wilson lines on $T^2$ for a $K3\times T^2$ compactification, is allowed by space-time supersymmetry. The main 
objective of the present work is to include them in the torsion GLSM and in the computation of the new supersymmetric index.

A recent impetus for studying superconformal sigma-models of K3 was the discovery of the 
{\it Mathieu moonshine}~\cite{Eguchi:2010ej}, which links such superconformal field theories, 
through the expansion of their $(2,2)$ elliptic genus into $\mathcal{N}=4$ characters, with representations of 
the sporadic group $\mathbb{M}_{24}$. Universality of $\mathcal{N}=2$ threshold 
corrections~\cite{Kiritsis:1996dn}, a consequence of 
their modular properties, implies that the same can be said for $K3\times T^2$ heterotic 
compactifications with arbitrary gauge bundle, $i.e.$ even 
for $(0,2)$ models that are not deformations of the standard embedding~\cite{2013JHEP...09..030C}. 

A natural question  is whether the Mathieu moonshine, or another type of moonshine, shows up in some form for the 
$\mathcal{N}=2$ compactifications with torsion studied here, that encompass the $K3\times T^2$ case. 
As a first step, we show that the dressed elliptic genus, that is the building block 
of the supersymmetric index,  admits a decomposition in terms of 
$\mathcal{N}=4$ superconformal characters for arbitrary consistent gauge bundle. In passing, 
we derive a congruence identity satisfied by the dimensions of $\mathbb{M}_{24}$ representations 
appearing in the moonshine module that is of general interest. Possible evidence for moonshine 
phenomena will be reported elsewhere~\cite{HIPS}. 

Finally, the general expression for the dressed elliptic genus in terms of standard 
weak Jacobi forms that we reach before the decomposition into 
$\mathcal{N}=4$  characters is a natural starting point for computing gauge and gravitational threshold corrections 
to the low energy effective action; these results will be presented in a forthcoming publication~\cite{AIS}. 

This article is organized as follows. In~\cref{sec:GLSM} we give a brief review of the torsional geometry 
of interest, describe the corresponding gauged linear sigma model and 
extend the formalism in order to include Abelian bundles.  In section~\ref{sec:index} we define its dressed elliptic genus and proceed to its computation using supersymmetric localization. 
The geometrical definition of this dressed elliptic genus in terms of a 
modified holomorphic Euler characteristic is introduced in \cref{sec:GeomForm}, and is shown to agree 
with the field theory definition for a large class of varieties in~\cref{app:ProofGeom}. 
The decomposition into $\mathcal N=4$ superconformal characters is discussed in~\cref{sec:ExpansionIndex}. 
$(0,2)$ superspace conventions are given in~\cref{app:Superspace}. 
Appendix~\ref{app:rat} focuses on the relation between the rational two-torus lattice 
and the rank two lattice defining the principal bundle. 

\paragraph{Conventions:}$\alpha'=1$. The action is written as
$S=\frac{1}{\pi}\int_\Sigma\text{d}^2 w\, \mathcal{L}$. 
The area of the worldsheet torus is $\int_\Sigma\text{d}^2 w=2\tau_2$. 
Left-moving corresponds to holomorphic in $w$. $T$ is the complex structure of the target-space 
torus, and $U$ its complexified K\"ahler modulus. 


\section{Torsional geometry and its gauged linear sigma-model}
\label{sec:GLSM}

We provide in this section a brief summary of the torsional geometry of interest, given by a principal two-torus 
bundle over a K3 surface, and  summarize the construction of the $(0,2)$ GLSM with 
torsion that describes the corresponding heterotic compactification. We will then show how to extend 
the formalism in order to include space-time Abelian gauge bundles over the total space of the principal torus bundle. 

\subsection{Heterotic $\mathcal{N}=2$ compactifications with torsion}

The geometry consists of a principal two-torus bundle over a warped $K3$ surface 
$\mathcal{S}$, $T^2\hookrightarrow X\stackrel{\pi}{\to}\mathcal{S}$. The metric follows from the ansatz:
\begin{equation}
\label{eq:metric}
\text{d}s^2=e^{2\Delta(y)}\d s^2(\mathcal{S})+\frac{U_2}{T_2}\left|\d x^1+T\d x^2+\pi^\star\alpha\right|^2\, ,
\end{equation}
where the warp factor $\Delta(y)$ depends only on the coordinates on $\mathcal{S}$, $\d s^2(\mathcal{S})$ is a 
Ricci-flat metric on  $\mathcal S$ and the complex connection one-form 
$\alpha=\alpha_1+T\alpha_2$  is such that $\iota:=\d x^1+T\d x^2+\pi^\star\alpha$ is a 
globally defined $(1,0)$-form on the total space $X$. 

$\mathcal{N}=1$ supersymmetry imposes some constraints on the curvature 
two-form $\omega = \omega_1 + T \omega_2$ in $\bigwedge^2 T^*_\mathcal{S}$, 
defined by $\tfrac{1}{2\pi}\, d \iota =   \pi^\star \omega$:
\begin{itemize}
\item $\omega$ has no $\bigwedge^{0,2}T^*_\mathcal{S}$ component,
\item $\omega$ is primitive with respect to the base, i.e. $\omega\wedge J_{\mathcal{S}}=0$.  
\end{itemize}
Furthermore, both $\omega_1$ and $\omega_2$ should belong to $H^2(\mathcal{S},\mathbb{Z})$, in order to get a well-defined 
bundle. 

Imposing that $\omega\in \bigwedge^{1,1}T^*_\mathcal{S}$ enhances space-time supersymmetry to $\mathcal{N}=2$. 
A further mild restriction is to consider that  both $\omega_1$ and $\omega_2$ are anti-self-dual $(1,1)$-forms. 
The bundle is then characterized by a pair of cohomology classes $[\omega_\ell]$ defining a sublattice of the Picard lattice of the base 
$\text{Pic}(\mathcal{S})=H^2(\mathcal{S},\mathbb{Z})\cap H^{1,1}_{\bar\partial}(\mathcal{S})$. 

A large class of vector bundles compatible with supersymmetry~\cite{Becker:2006et} consists first of the pullback of 
a stable holomorphic vector bundle over $\mathcal{S}$, satisfying the integrated Bianchi identity:
\begin{equation}
\label{eq:tad}
\int_\mathcal{S}\text{ch}_2(\mathcal{V})+24-\frac{U_2}{T_2}
\int_\mathcal{S}\omega\wedge\star_{\scriptscriptstyle{\mathcal{S}}}\bar\omega=0\, .
\end{equation}
We assume in the following that the structure group of this vector bundle is embedded in the 
first $E_8$ factor of the heterotic gauge group. 

Second, one can consider also an Abelian bundle over the total space $X$, whose connection is of the form:
\begin{equation}
\label{eq:connection}
A=\mathcal{T}^a\, \text{Re}(\bar V^a\, \iota)\, ,
\end{equation}
depending on 8 complex parameters $V^a$. It reduces to a set of 
Wilson lines for $K3\times T^2$ compactifications, which constitute particular cases of this construction. 
Therefore, we will loosely call them Wilson lines thereafter. 

For simplicity we will embed the structure group of this bundle in the second $E_8$ factor. In eq.~(\ref{eq:connection}) 
$\{\mathcal{T}^a\}$ forms a basis of $\mathfrak{H}_8$, its Cartan subalgebra. The 
tadpole condition~(\ref{eq:tad}) is unchanged hence depends only on the second Chern character 
of the vector bundle over the base and on the torus moduli $T$ and $U$. The present paper extends the results 
of~\cite{Israel:2015aea} where only special points 
in the moduli space where these Abelian gauge bundles over the total space where turned off were considered. 

\subsection{The gauged linear sigma-model with torsion}

We briefly review  the construction of GLSMs with torsion introduced in~\cite{Adams:2006kb}.

\subsubsection{The \texorpdfstring{$K3$}{K3} base}

We refer the reader to~\cref{app:Superspace} for $(0,2)$ superspace conventions; the worldsheet
 gauge group is taken to be $U(1)$ here for clarity of the presentation. 
The $K3$ base $\mathcal{S}$ corresponds to a standard $(0,2)$ 
GLSM (see~\cite{1995hepc.conf..322D} for a review) with chiral multiplets $\{\Phi_I\}_{I=0,\dots,n}$ of positive 
charge, $P$ of negative charge and two sets of Fermi multiplets, 
$\{\Gamma_a\}_{\alpha=0,\dots,r}$ of positive charge and  $\{\tilde\Gamma_\alpha\}_{\alpha=1,\dots,p}$ of 
negative charge. We denote their gauge charges by 
$Q_I$, $Q_P$, $Q_\alpha$ and $Q_a$ respectively. The superpotential is composed of 
two pieces. First, 
\begin{equation}
\mathcal{L}_{\mathcal{S}}=\int\text{d}\theta\, \tilde\Gamma_\alpha G^\alpha(\Phi) +\text{h.c.}\, ,
\end{equation}
where $G^\alpha$ are quasi-homogeneous polynomials and define a codimension $p$ subvariety 
$\mathcal{S}$ in an ambient $n$-dimensional weighted projective space as the complete intersection:
\begin{equation}
\bigcap_{\alpha=1}^{p}\left\{\phi_I|G^\alpha(\phi_I)=0\right\}\,  ,
\end{equation}
that is chosen to obey the Calabi-Yau condition. The second piece,
\begin{equation}
\label{eq:pterm}
\mathcal{L}_{\mathcal{V}}=\int\text{d}\theta\, P\Gamma_a J^a(\Phi)+\text{h.c.}\, ,
\end{equation}
where $J^a$ are again quasi-homogeneous polynomials defines a monad gauge bundle $\mathcal{V}$ of rank $r$ through the following short exact sequence:
\begin{equation}
0\rightarrow \mathcal{V}\rightarrow \bigoplus_{a=0}^{r}\mathcal{O}(Q_a)\overset{\otimes J^a}{\rightarrow}\mathcal{O}(-Q_P)\rightarrow 0\, .
\end{equation}
The model contains 
chiral fermions hence is potentially anomalous, the variation of the effective Lagrangian under a super-gauge transformation of chiral parameter $\Xi$ being:
\begin{equation}
\label{eq:BaseAnomaly}
\delta_\Xi \mathcal{L}_{\text{eff}}=-\frac{\mathcal{A}}{8}\int\text{d}\theta\, \Xi\Upsilon+\text{h.c.}\, ,
\end{equation}
with \begin{equation}
\mathcal{A}=\sum_{\text{chiral}}Q^2-\sum_{\text{fermi}}Q^2\, .
\end{equation}
Allowing for a non-vanishing anomaly, the model is at this point ill-defined quantum mechanically.

\subsubsection{The torus fiber}
We consider a generic two-torus of metric and B-field:
\begin{equation}
\label{eq:torusdata}
g=\frac{U_2}{T_2}\begin{pmatrix}
1&T_1\\T_1&|T|^2
\end{pmatrix}\, ,\ \ \ \ \ b=\begin{pmatrix}
0&U_1\\-U_1&0
\end{pmatrix}\, ,
\end{equation}
and introduce two extra chiral superfields $\{\Omega^\ell=(\omega^\ell,\chi^\ell)\}_{\ell=1,2}$, which are charged axially under the super-gauge symmetry:
\begin{equation}
\delta_\Xi(\Omega^\ell)=im^\ell\, \Xi\ , \quad m^\ell \in \mathbb{Z} \, .
\end{equation}
For this reason we will call them 'shift multiplets' in the following. For a torus of metric and 
B-field~(\ref{eq:torusdata}) one considers the following Lagrangian density:
\begin{align}
\mathcal{L}^0_{\text{tor}}=&-\frac{iU_{2}}{8T_{2}}\int\text{d}^2\theta\ \Big(\Omega^{1}+\bar{\Omega}^{1}+T_{1}\left(\Omega^{2}+\bar{\Omega}^{2}\right)
+2(m^1+T_{1}m^2)\mathcal{A}\Big)\times\nonumber\\
&\times\Big(\partial_{-}\left(\Omega^{1}-\bar{\Omega}^{1}+T_{1}\left(\Omega^{2}-\bar{\Omega}^{2}\right)\right)
+2i(m^1+T_{1}m^2)\mathcal{V}\Big)\nonumber\\
&-\frac{iU_{2}T_{2}}{8}\int\text{d}^2\theta\ \Big(\Omega^{2}+\bar{\Omega}^{2}+2m^2\mathcal{A}\Big)
\Big(\partial_{-}\left(\Omega^{2}-\bar{\Omega}^{2}\right)+2im^2\mathcal{V}\Big)\nonumber\\
&+\frac{iU_{1}}{8}\int\text{d}^2\theta\ \Big\{\Big(\Omega^{1}+\bar{\Omega}^{1}+2m^1\mathcal{A}\Big)
\Big(\partial_{-}\left(\Omega^{2}-\bar{\Omega}^{2}\right)+2im^2\mathcal{V}\Big)\nonumber\\
&-\Big(\Omega^{2}+\bar{\Omega}^{2}+2m^2\mathcal{A}\Big)\Big(\partial_{-}\left(\Omega^{1}-
\bar{\Omega}^{1}\right)+2im^1\mathcal{V}\Big)\Big\}\nonumber\\
&-\frac{ih_\ell}{4}\int\text{d}\theta\, \Upsilon\, \Omega^\ell+\text{h.c.}\, ,
\end{align}
in which the shift multiplets are coupled to the gauge field both minimally, and axially via a field-dependent 
Fayet-Iliopoulos term (last line). Their imaginary part will eventually model the two-torus fiber. 

\subsubsection{Anomaly cancellation and moduli quantization}

The key point is that the above Lagrangian is classically not invariant under a super-gauge transformation of 
chiral superfield parameter $\Xi$, but rather transforms as:
\begin{equation}
\delta_\Xi(\mathcal{L}^0_{\text{tor}})=\frac{h_\ell m^\ell}{4}\int\text{d}\theta\, \Upsilon\Xi +\text{h.c.}\, ,
\end{equation}
due to the field-dependent Fayet-Iliopoulos coupling. One can use this classical variation to cancel the gauge anomaly 
coming from the K3 base, see~\cref{eq:BaseAnomaly}. The non-vanishing torsion flux 
is therefore implemented at the GLSM level as a two-dimensional Green-Schwarz mechanism~\cite{Adams:2006kb}. A consistent 
model should also contain a non-anomalous global right-moving $U(1)_\textsc{r}$ symmetry flowing in the IR to the 
R-symmetry, and a left-moving $U(1)_\textsc{l}$ used to define a left spectral flow.

At this point, the geometry obtained after integrating out the massive gauge field is 
that of a $\left(\mathbb{C}^*\right)^2$ bundle over $K3$. To decouple the real part of 
the shift multiplets $\Omega^\ell$, in order to restrict to a $T^2$ bundle while preserving 
$(0,2)$ supersymmetry, one  should cancel their couplings to the gaugini. This leads to the following conditions:
\begin{subequations}
\label{eq:quant}
\begin{align}
\frac{U_2}{T_2}(m^1+T_1m^2)-U_1m^2+h_1&=0\\
\frac{U_2}{T_2}\left[(m^1+T_1m^2)T_1+T_2^2m^2\right]+U_1m^1+h_2&=0\, .
\end{align}
\end{subequations}
Demanding that the action is single-valued in every topological sector imposes $h_\ell\in\mathbb{Z}$, 
giving restrictions on the torus moduli. With 
at least a rank-two worldsheet gauge group, $U$ and $T$ are generically quantized such that the 
underlying $c=2$ CFT with a 2-torus target space is rational, see app.~\ref{app:rat}. In the 
case of a $U(1)$ worldsheet gauge group, as discussed here for simplicity of the presentation, 
one complex torus modulus remains unfixed. 

Using the relations~(\ref{eq:quant}), the anomaly cancellation condition can be written in a simple form:
\begin{equation}
\label{eq:Anom}
\mathcal{A}-\frac{2U_2}{T_2}|\mathfrak{m}|^2=0\, ,
\end{equation}
where $\mathfrak{m}:=m^1+Tm^2$ is the complex topological charge, which is the worldsheet counterpart of the 
tadpole condition~(\ref{eq:tad}).

After the decoupling of the real part of $\Omega^\ell$ is done, one can rearrange the remaining degrees of freedom into 
a 'torsion multiplet'~\cite{2011JHEP...03..045A}, in order to exhibit more explicitly the torus sub-bundle 
inside the $\left(\mathbb{C}^*\right)^2$ bundle. 
That approach was adopted in~\cite{Israel:2015aea}, but we will stick here to a formulation in terms of shift multiplets as 
the Abelian bundle in target-space will be more naturally described in this framework. 

\subsection{Abelian connections over the total space}

In order to describe a target-space Abelian gauge bundle over the total space $X$ as~(\ref{eq:connection}), 
one needs to enlarge the torsion GLSM framework. For simplicity, we embed the structure group of $\mathcal{V}$ in 
the first $E_8$ and the structure group of the Abelian bundle in the second $E_8$.

From the worldsheet perspective, each line bundle is mapped to a left-moving Weyl fermion $\lambda_-$ 
in a Fermi multiplet $\Lambda$, transforming as a section of this bundle. 
In  components, a connection of the type (\ref{eq:connection}) corresponds to a kinetic term like 
$\bar\lambda_-\left(\partial_+\left(\omega-\bar\omega\right)+2mA_+\right)\lambda_-$ in the 
Lagrangian of the two-dimensional supersymmetric gauge theory.  It will be convenient to 
bosonize these left-moving fermions, as one will be able to consider them and the shift multiplets for the two-torus 
on the same footing.

As one defines the GLSM in $(0,2)$ superspace, one needs to add enough degrees of freedom to form a multiplet. 
One first bosonizes $\lambda_-$ into a chiral and 
real compact boson,  and embeds it in a neutral chiral multiplet $B$, of components $B=(b,\bar b,\xi_+,\bar \xi_+)$, 
as the left-moving, compact imaginary part of $b$. Of course, such a procedure 
introduces extra degrees of freedom. For each multiplet $B$, one has:
\begin{itemize}
\item The real part of $b$ which is non-compact,
\item The right-moving fermions $\xi_+$ and $\bar\xi_+$,
\item The right-moving part of $\text{Im}(b)$.
\end{itemize}
All of these extra degrees of freedom are an artifact of the bosonization procedure. Naturally the right-moving part of 
$\text{Im}(b)$ cannot decouple from the Lagrangian of the theory, as it would give Lagrangians 
for chiral bosons. However as we shall see at the end of the computation, the contribution from those degrees of freedom 
will appear in the dressed elliptic genus as an overall finite and non-vanishing multiplicative factor.

The dynamics of the chiral multiplets $\{B^n\}_{n=1,\dots,8}$ is described by the following Lagrangian:
\begin{align}
\label{eq:bnlag}
\mathcal{L}_{\text{Wilson}}=&-\frac{i\mathcal{E}_{mn}}{8}\int \text{d}^2\theta\, 
\Big(B^m+\bar B^m\Big)\partial_-\left(B^n-\bar B^n\right)\nonumber\\
&-\frac{i\beta_{\ell n}}{16}\int \text{d}^2\theta\, \Big(\Omega^\ell+\bar\Omega^\ell+2m^\ell\mathcal{A}\Big)\partial_-\left(B^n-\bar B^n\right)\, ,
\end{align}
where $\mathcal{E}_{mn} : = \mathcal{G}_{mn} + \mathcal{B}_{mn}$ is such that the corresponding $(8,8)$ toroidal lattice 
splits into $(E_8)_\textsc{l} \times (E_8)_\textsc{r}$, $i.e.$ 
into two lattices of signatures $(8,0)$ and $(0,8)$ respectively, both 
isomorphic to the $E_8$ root lattice, see $e.g.$~\cite{Elitzur:1986ye}. 

The $B^n$'s are chirally coupled to the torus shift multiplets through the off-diagonal terms in the second line of eq.~(\ref{eq:bnlag}), 
leading to couplings corresponding to the connection~\cref{eq:connection} in space-time.  
The parameters $\beta_{\ell n}$ are related to the 'Wilson line' moduli $V^a$, see eqs.~(\ref{eq:vbeta}) in the next section.  
Unlike the torus moduli $(T,U)$, they are not quantized by the flux. 

A discussion about moduli quantization in this context, from the target-space viewpoint, can be found 
in~\cite{Melnikov:2012cv}. In that article examples where the Abelian bundle was not embeded in the commutant 
of the structure group of $\mathcal V$ were also considered. They can be 
incorporated in the present framework without too much effort. One needs to gauge the imaginary shift symmetry of the 
$B^n$'s, and add an extra axial coupling of the form  
$\int\text{d}\theta\, \Upsilon\, B^n$, in order to reproduce the gauge anomaly; in other words, 
at least part of the $B^n$'s become shift multiplets similar 
to the $\Omega^\ell$'s modeling the two-torus fiber.

\subsubsection*{The extended fiber Lagrangian}

In the following, we will adopt compact notations incorporating both the torus  and the 'Wilson lines' by 
working with a $(10,10)$ lattice whose metric and B-field are:
\begin{equation}
\label{eq:extlattice}
G:=\begin{pmatrix}[cc|cccc]
\frac{U_2}{T_2}&\frac{U_2}{T_2}T_1&\frac{\beta_{11}}{4}&\frac{\beta_{12}}{4}&\cdots&\frac{\beta_{18}}{4}\\[1mm]
\frac{U_2}{T_2}T_1&\frac{U_2}{T_2}|T|^2&\frac{\beta_{21}}{4}&\frac{\beta_{22}}{4}&\cdots&\frac{\beta_{28}}{4}\\[1mm] \hline
\rule{0pt}{2.5ex}
\frac{\beta_{11}}{4}&\frac{\beta_{21}}{4}&\mathcal{G}_{11}&\mathcal{G}_{12}&\cdots&\mathcal{G}_{18}\\[1mm]
\frac{\beta_{12}}{4}&\frac{\beta_{22}}{4}&\mathcal{G}_{21}&\mathcal{G}_{22}&\cdots&\mathcal{G}_{28}\\[1mm]
\vdots&\vdots&\vdots&\vdots&\ddots&\vdots\\[1mm]
\frac{\beta_{18}}{4}&\frac{\beta_{28}}{4}&\mathcal{G}_{81}&\mathcal{G}_{82}&\cdots&\mathcal{G}_{88}\\
\end{pmatrix}
,\ \ \ \ \ 
B:=\begin{pmatrix}[cc|cccc]
0&U_1&\frac{\beta_{11}}{4}&\frac{\beta_{12}}{4}&\cdots&\frac{\beta_{18}}{4}\\[1mm]
U_1&0&\frac{\beta_{21}}{4}&\frac{\beta_{22}}{4}&\cdots&\frac{\beta_{28}}{4}\\[1mm] \hline
\rule{0pt}{2.5ex}
-\frac{\beta_{11}}{4}&-\frac{\beta_{21}}{4}&0&\mathcal{B}_{12}&\cdots&\mathcal{B}_{18}\\[1mm]
-\frac{\beta_{12}}{4}&-\frac{\beta_{22}}{4}&\mathcal{B}_{21}&0&\cdots&\mathcal{B}_{28}\\[1mm]
\vdots&\vdots&\vdots&\vdots&\ddots&\vdots \\[1mm]
-\frac{\beta_{18}}{4}&-\frac{\beta_{28}}{4}&\mathcal{B}_{81}&\mathcal{B}_{82}&\cdots&0\\
\end{pmatrix}\, .
\end{equation}
We also introduce the following combinations:
\begin{equation}
E=G+B\, ,\ \ \ \ \bar E=G-B .
\end{equation}

Let us group together the gauge charges and Fayet-Iliopoulos couplings into the following vectors, and denote the 
various (shift) multiplets by a common letter:
\begin{equation}
v:=\begin{pmatrix}
m^1\\m^2\\ \hline 0\\ \vdots \\ \vdots \\ 0
\end{pmatrix}
,\ \ \ \ \ 
h:=\begin{pmatrix}
h_1\\h_2\\ \hline 0\\ \vdots \\ \vdots \\ 0
\end{pmatrix}
,\ \ \ \ \ 
\Omega:=\begin{pmatrix}
\Omega^1\\\Omega^2\\ \hline B^1\\B^2\\ \vdots \\ B^8
\end{pmatrix}\, .
\end{equation}
The indices $\{i,j\}$ run over the full set of multiplets $\{\Omega^i\}_{i=1,\dots,10}$ thereafter. 

With these notations, the Lagrangian $\mathcal{L_\text{tor}}=\mathcal{L}^0_\text{tor}+\mathcal{L}_\text{Wilson}$ modelling the two-torus together with a set of $8$ complex Wilson lines reads:
\begin{align}
\label{eq:torusLag}
\mathcal{L_\text{tor}}=&-\frac{iE_{ij}}{8}\int \text{d}^2\theta\, \Big(\Omega^i+\bar \Omega^i+2v^i\mathcal{A}\Big)\Big(\partial_-\left(\Omega^j-\bar \Omega^j\right)+2iv^j \mathcal{V}\Big)\nonumber\\
&-\frac{ih_i}{4}\int\text{d}\theta\, \Upsilon \Omega^i+\text{h.c.}\, .
\end{align}
Upon using the conditions~\eqref{eq:quant}, the Lagrangian~\eqref{eq:torusLag} is given in components, after integrating by parts by:
\begin{align}
\mathcal{L}=\, &\frac{E_{ij}}{8}\Big\{\partial_+\left(\omega^i+\bar\omega^i\right)\partial_-\left(\omega^j+\bar\omega^j\right)-\partial_+\left(\omega^i-\bar\omega^i\right)\partial_-\left(\omega^j-\bar\omega^j\right)\nonumber\\
&-2iv^i\partial_-\left(\omega^j-\bar\omega^j\right)A_+-2iv^j\partial_-\left(\omega^i-\bar\omega^i\right)A_++4v^iv^jA_+A_-\nonumber\\
&+2i\chi^i\partial_-\bar\chi^j+2i\bar\chi^i\partial_-\chi^j\Big\} + \text{t.d.}\, .
\label{eq:ShiftLagrangian}
\end{align}


\section{Dressed elliptic genus of compactifications with torsion}
\label{sec:index}

In this section we will define the dressed elliptic genus of the torsion GLSM, and obtain its 
expression using supersymmetric localization.

\subsection{Dressed elliptic genus and new supersymmetric index}

A natural supersymmetric index of two-dimensional superconformal field theories with $(0,2)$ supersymmetry 
and a (non-anomalous) global $U(1)_\textsc{l}$ symmetry, in particular non-linear sigma models on Calabi-Yau $n$-folds with a holomorphic vector 
bundle, is given by their elliptic genus~\cite{Witten:1986bf,Alvarez:1987de,Alvarez:1987wg,Kawai:1993jk,Kawai:1994np}, defined in the Hamiltonian 
formalism as the partition function with periodic boundary conditions:
\begin{equation}
Z_{\text{Ell}}\left(\tau,z\right) = \text{Tr}_{\textsc{rr}}\left\{  e^{2i\pi z J_{0}}
(-1)^{F} q^{L_0-c/24}\bar{q}^{\bar{L}_0-\bar{c}/24} \right\}\, , 
\end{equation}
corresponding to the Witten index with a chemical potential for the zero-mode $J_{0}$ of the left-moving  $U(1)_\textsc{l}$ 
current inserted. 
This elliptic genus is also defined mathematically as the holomorphic Euler characteristic of some formal power 
series with bundle coefficients, see section~\ref{sec:GeomForm}. 

The elliptic genus vanishes identically for the $\mathcal N = 2$ torsional compactifications 
of interest, because of the right-moving fermionic zero-modes of the torsion multiplet that cannot be saturated 
(as they do not appear in the interactions); the same holds for ordinary 
$K3 \times T^2$ compactifications.

The new supersymmetric index~\cite{1992NuPhB.386..405C} is the natural non-vanishing supersymmetric index appearing 
in the context of $\mathcal{N}=2$ heterotic compactifications, for instance when computing 
threshold corrections to the gauge couplings~\cite{1996NuPhB.463..315H}. 
It is defined by the following trace in the Ramond sector of the right-moving fermions:
\begin{equation}
Z_\textsc{new}(\tau,\bar\tau)=\frac{1}{\eta(\tau)^2}\text{Tr}_\textsc{r}
\left\{  \bar{J}_0 (-1)^{F_R} q^{L_0-c/24}\bar{q}^{\bar{L}_0-\bar{c}/24} \right\}\, .\end{equation}
For $K3 \times T^2$ it can easily be expressed in terms of the $K3$ elliptic genus. 

\subsubsection{Compactifications without Abelian bundles over the total space}

In the context of $\mathcal N=2$ compactifications with torsion, without 'Wilson lines' for the moment, 
let us first define a modified elliptic genus appropriate to the torsional compactifications 
of interest, which will eventually correspond to the (anomalous) elliptic genus of the $K3$ base dressed by the 
non-holomorphic contribution of the two-torus fiber. Explicitly, this {\it dressed elliptic genus}, 
introduced in~\cite{Israel:2015aea}, 
is defined as:
\begin{equation}
Z^0_\textsc{fy}\left(\tau,\bar{\tau},z\right) = \frac{1}{\bar{\eta}(\bar{\tau})^{2}}\, 
\text{Tr}_{\mathcal{H}^0,\textsc{rr}}\left\{  e^{2i\pi z J_{0}}\bar{J}_{\, 0} 
(-1)^{F} q^{L_0-c/24}\bar{q}^{\bar{L}_0-\bar{c}/24} \right\}\, ,
\label{eq:FYpartzero}
\end{equation}		
the trace being taken in the Hilbert space $\mathcal{H}^0$ of the $(0,2)$ superconformal sigma-model on 
$T^2\hookrightarrow X\stackrel{\pi}{\to}\mathcal{S}$, in the Ramond--Ramond sector. This object corresponds 
therefore to the elliptic genus with an extra  insertion of the zero-mode $\bar J_0$ of the right-moving 
R-symmetry $U(1)_R$ current, and an overall $1/\bar\eta^2$ factor added for later  
convenience. For a rank $r$ vector bundle, embedded into the 
first $E_8$ factor, the dressed elliptic genus 
gives the new supersymmetric index after performing a left-moving GSO projection as follows:
\begin{equation}
Z_\textsc{new}(\tau,\bar\tau)=\frac{\bar{\eta}^{2}E_{4}(\tau)}{2\eta^{10}}\sum_{\gamma,\delta=0}^{1}q^{\gamma^{2}}\left.
\left\{\left(\frac{\vartheta_{1}\left(\tau\left|z\right.\right)}{\eta(\tau)}\right)^{8-r}
Z^0_\textsc{fy}\left(\tau,\bar{\tau},z\right)\right\}\right|_{z=\frac{\gamma\tau+\delta}{2}}\, .
\end{equation}

\subsubsection{Compactifications with Abelian bundles over the total space}

In the formulation of the GLSM used in this work, unlike in our previous article~\cite{Israel:2015aea}, one has also 
to deal with the spurious degrees of originating from the shift multiplets $\Omega^\ell$ and 
from the $B^n$ multiplets. All these degrees of freedom are of course artifacts of this 
formulation and should be decoupled at the end of the computation. 

As an intermediate step, one defines a supersymmetric index appropriate for this 'enlarged' $(0,2)$ superconformal 
field theory as follows:
\begin{equation}
Z_\textsc{ext}\left(\tau,\bar{\tau},z\right) = \frac{1}{\bar{\eta}(\bar{\tau})^{20}}\, 
\text{Tr}_{\mathcal{H}^\textsc{ext},\textsc{rr}}\left\{  e^{2i\pi z J_{0}}\left(\bar{J}_{\, 0}\right)^{10}
(-1)^{F} q^{L_0-c/24}\bar{q}^{\bar{L}_0-\bar{c}/24} \right\}\, ,
\label{eq:FYpartint}
\end{equation}		
the trace being taken in the Hilbert space $\mathcal{H}^\textsc{ext}$ of the 
SCFT at the infrared fixed point of the torsion GLSM comprising the 
shift multiplets $\{\Omega^i\}_{i=1,\dots,10}$, in the left and right Ramond sectors.

The extra insertions of the R-current zero mode $\bar J_0$ in~(\ref{eq:FYpartint}) are needed in order to cancel the 
extra spurious fermionic zero modes appearing in this formulation. 
The right-moving R-current of the $(0,2)$ GLSM with the multiplets $\{\Omega^i\}_{i=1,\dots,10}$ is:
\begin{equation}
\bar J=G_{ij}\, \bar\chi^i\chi^j+\ldots\, ,
\end{equation}
where the ellipsis stands for the contribution of the other fields of the theory. From the path integral point of view, 
this means that each $\bar J_0$ insertion has indeed the effect of saturating the fermionic zero modes of a 
fermion contained in a shift multiplet; hence, having inserted just the right power of this zero-mode, 
the other terms contributions to the current $\bar J_R$ do not play any role in the computation.  
Additionally to the right-moving fermions $\chi^i$, one 
gets first a  contribution from the non-compact real part of the bosons $\omega^i$, which is completely factorized. 
Remains finally the contributions from the right-moving part of $\text{Im}(b^n)$, which will be discussed in due time.

From this intermediate partition function $Z_\textsc{ext}$  one can then extract the 
dressed elliptic genus of interest that we define as, 
\begin{equation}
Z^\textsc{w}_\textsc{fy}\left(\tau,\bar{\tau},z\right) = \frac{1}{\bar{\eta}(\bar{\tau})^{2}}\, 
\text{Tr}_{\mathcal{H}^\textsc{w}_\textsc{rr}}\left\{  e^{2i\pi z J_{0}}\bar{J}_{\, 0} 
(-1)^{F} q^{L_0-c/24}\bar{q}^{\bar{L}_0-\bar{c}/24} \right\}\, ,
\label{eq:FYpart}
\end{equation}		
where $\mathcal{H}^\textsc{w}_\textsc{rr}$ is the Hilbert space of the SCFT corresponding to the $(0,2)$ 
non-linear sigma model of central charges $(c,\bar c) = (14+r,9)$  and target space 
$T^2\hookrightarrow X\stackrel{\pi}{\to}\mathcal{S}$, with a rank $r$ gauge bundle $\mathcal{V}$ 
in the first $E_8$ factor and a generic Abelian gauge bundle in the second $E_8$ factor; while the trace is 
restricted to the left Ramond sector for the former, we sum over all spin structures for the latter. 

The index~(\ref{eq:FYpart}) is the closest analogue of the elliptic genus in the present context, 
and consists in a non-holomorphic dressing of the elliptic genus of the $K3$ base, 
which is anomalous with respect to modular transformations, by a $(10,2)$ lattice encoding 
the principal two-torus bundle and the line bundles over its total space. The new supersymmetric 
index is then obtained as
\begin{equation}
Z_\textsc{new}(\tau,\bar\tau)=\frac{\bar{\eta}^{2}}{2\eta^{2}}\sum_{\gamma,\delta=0}^{1}q^{\gamma^{2}}\left.
\left\{\left(\frac{\vartheta_{1}\left(\tau\left|z\right.\right)}{\eta(\tau)}\right)^{8-r}
Z^\textsc{w}_\textsc{fy}\left(\tau,\bar{\tau},z\right)\right\}\right|_{z=\frac{\gamma\tau+\delta}{2}}\, .
\end{equation}

\subsection{Computation of the dressed elliptic genus through localization}

The supersymmetric partition function~(\ref{eq:FYpartint}) corresponds, in Lagrangian formalism, to the following 
path integral on an Euclidean torus of complex structure $\tau$:
\begin{align}
\label{eq:FYcompact}
Z_\textsc{ext} (\tau,\bar \tau, z) =&\, \frac{1}{\bar{\eta}(\bar{\tau})^{20}}\int \mathscr{D} a_{w} \mathscr{D} a_{\bar w}\mathscr{D} \lambda \mathscr{D} 
\bar{\lambda} \mathscr{D} D \, 
e^{-\frac{1}{e^2} S_{\text{Gauge}}[a, \lambda, D]- t\, S_{\textsc{fi}} (a,D)}\ \times\nonumber \\ &\times \  
\int \prod_I\mathscr{D} \phi_{I}\mathscr{D} \bar{\phi}_{I}  \mathscr{D} \psi_{I} \mathscr{D} \bar{\psi}_{I}  \, 
e^{-\frac{1}{g^2} S_{\text{chiral}}[\phi_I,\psi_I,a,D,a_\textsc{l}]}
\ \times\nonumber \\ &\times \ 
\int \prod_a\mathscr{D} \gamma_{a}\mathscr{D} \bar{\gamma}_{a}  \mathscr{D} G_a \mathscr{D} \bar{G}_{a}  \, 
e^{-\frac{1}{f^2} 
S_{\text{Fermi}}[\gamma_{a},G_a,a,a_\textsc{l}]-S_{\text{pot}}[\gamma_{a},G_a,\phi_i,\psi_{i}]}
\ \times\nonumber \\ &\times \ 
\int \prod_{i=1}^{s+2}\mathscr{D} \omega^i \mathscr{D} \bar{\omega}^i  \mathscr{D} \chi_i \mathscr{D} \bar{\chi}_i  \, 
e^{-S_{\text{tor}}[\omega^i,\chi_i,a,a_\textsc{l}]}\left(\int \frac{\d^2 w}{2\tau_2} \,  
G_{ij}\, \bar\chi_i\chi_j\right)^{10}\, ,
\end{align}
where we have included couplings to a background gauge field for the $U(1)_{\textsc{l}}$ global symmetry
\begin{equation}
\label{eq:backgd}
a_\textsc{l} = \frac{\pi z}{i \tau_2}\, {\rm d}w-\frac{\pi z}{i \tau_2}\, {\rm d}\bar w\, ,
\end{equation}
in order to implement the twisted boundary conditions, as well as coupling constants $g$ and $f$ in front 
of the chiral and Fermi Lagrangians for convenience.

Following~\cite{2014LMaPh.104..465B} and~\cite{Israel:2015aea},  
this path integral localizes to the BPS configurations with respect to the supercharge:
\begin{equation}
\mathcal{Q}=\left.\left(\epsilon \mathcal{Q}_{+}-\bar{\epsilon}\bar{\mathcal{Q}}_{+}+\delta_{\textsc{wz}}\right)\right|_{\epsilon=\bar\epsilon=1}\, ,
\end{equation}
with $\delta_{\textsc{wz}}$ the super-gauge transformation of chiral parameter 
$\Xi_{\textsc{wz}}=i\bar{\epsilon}\theta a_{\bar{w}}$ needed to restore Wess-Zumino gauge.

One can wonder whether standard localization arguments apply to the path integral~(\ref{eq:FYcompact}).  
Indeed, as was emphasized above, neither the action, because of the field dependent Fayet-Ilioupoulos couplings, 
nor the path integral measure, because of the gauge anomaly, are separately gauge-invariant hence  
supersymmetry-invariant. 
However, owing to the anomaly cancellation 
condition \cref{eq:Anom}, one has:
\begin{equation}
\mathcal{Q}\left(\mathscr{D}\Phi\mathscr{D}\Gamma\, e^{-S}\right)=0\, .
\end{equation}
Moreover, the operator $\int {\rm d}^2 w \, G_{ij}\, \bar\chi^i\chi^j$ is not annihilated by 
$\mathcal Q$. Thankfully, terms generated by the action of the supercharge 
do not saturate the fermionic measure hence do not contribute to the path integral.
Finally, one can show as in~\cite{2014LMaPh.104..465B} 
that the whole Lagrangian is actually $\mathcal{Q}$-exact, apart from the torus fiber part.

Gathering these arguments, one can see that the path integral does not depend on the various 
couplings of the theory, allowing to compute it in the free-field limit:
\begin{equation}
e,g,f\rightarrow 0\, .
\end{equation}
Notice that even though non-$\mathcal{Q}$-exact, the torus part is Gaussian hence can be computed directly. 
It implies as expected that the result will depend on the two-torus moduli $(T,U)$ as well as on the 
'Wilson lines' moduli $V^a$.
 
The localization procedure reduces the path integral to a finite-dimensional integral over 
the gauge holonomies $(u,\bar u)$ on the worldsheet torus, 
the zero-modes of the gauginos and of the auxiliary $D$-field. 
We refer to \cite{2014LMaPh.104..465B} and \cite{Israel:2015aea} for details and 
for the reduction of the final result to a contour integral in the $u$ 
complex plane of the one-loop determinant. In the following, we 
will just summarize the various contributions to this determinant, from the base and the torus fiber.
 
\subsubsection{Contribution from the $K3$ base}

The $K3$ base contribution is made of standard chiral and Fermi multiplets. They contribute to the integral via their 
one-loop determinant,  which involves choosing a prescription for the determinant of chiral Dirac operators. 
The prescription we choose is the following one:
\begin{equation}
\text{Det}\, \nabla(u)=e^{\frac{\pi}{\tau_{2}}(u^{2}-u\bar{u})}\vartheta_{1}(\tau|u)\, ,
\end{equation}
which is compatible with the contribution of the torus fiber, see later. With this prescription the one-loop determinants of the 
base fields are then:
\begin{subequations}
\begin{align}
\text{chiral:   }Z_{\Phi_{I}}(\tau,u,\tilde u,z)&=i\, e^{-\frac{\pi}{\tau_{2}}(\upsilon^{2}-\upsilon\bar{\upsilon})}\, 
\frac{\eta(\tau)}{\vartheta_{1}(\tau|\upsilon)},\ \ \ \ \upsilon = Q_I u + q^{\textsc{l}}_{I}z\, ,\\
\text{Fermi:   }Z_{\Gamma_{a}} (\tau,u,\tilde u,z) &=i\, e^{\frac{\pi}{\tau_{2}}(\upsilon^{2}-\upsilon\bar{\upsilon})}\, 
\frac{\vartheta_{1}(\tau|\upsilon)}{\eta(\tau)},\ \ \ \ \ \, \upsilon = Q_a u + q^{\textsc{l}}_{a}z\, ,
\end{align}
\end{subequations}
where $Q_i$ (resp. $q^\textsc{l}_i$) denotes the gauge charge (resp. the global $U(1)_\textsc{l}$ charge) of the multiplet. The global  
charges are chosen in such a way that the possible $U(1)_\textsc{l}$ anomalies cancel, see~\cite{Israel:2015aea} for details.

Finally the contribution of the vector multiplet, for a $U(1)$ gauge group, is given simply by:
\begin{equation}
Z_{A} (\tau) =-2i\pi\eta(\tau)^{2}\, .
\end{equation}

\subsubsection{Contribution from the extended fiber}

We compute below the contribution of the 'extended' fiber Lagrangian, introduced in section~\ref{sec:GLSM}, eq.~(\ref{eq:torusLag}), containing the chiral multiplets $\{\Omega^i\}_{i=1,\dots,10}$.

We consider first the bosonic terms in the Lagrangian~(\ref{eq:ShiftLagrangian}). Let us define the compact bosons:
\begin{equation}
\alpha^i:=\text{Im}(\omega^i)\, ,
\end{equation}
and proceed to a Wick rotation. $\alpha^1$ and $\alpha^2$ describe the coordinates on the two-torus of moduli 
$T$ an $U$, while the other $\alpha^i$ correspond to the lattice $(E_8)_\textsc{l}\times (E_8)_\textsc{r}$.
Setting aside the decoupled real part of $\omega^i$, the bosonic part of the Lagrangian is then:
\begin{equation}
\label{eq:bosLag}
\mathcal{L}_\text{bos}=\frac{E_{ij}}{2}\Big\{\bar\partial\alpha^i\partial\alpha^j+v^i\partial\alpha^j A_{\bar w}+v^j\partial\alpha^i A_{\bar w}+v^iv^jA_wA_{\bar w}\Big\}\, .
\end{equation}
The fields $\alpha^i$ satisfy the periodicity conditions:
\begin{equation}
\alpha^i(z+k+\tau l,\bar z+k+\bar\tau l)=\alpha^i(z,\bar z)+2\pi(kw_i+ln_i)\, .
\end{equation} 
The zero mode part of the compact bosons is then:
\begin{equation}
\alpha^i_0 (z,\bar z)=\frac{i\pi}{\tau_2}\left\{z(w_i\bar\tau-n_i)-\bar z(w_i\tau-n_i)\right\}\, ,
\end{equation}
where $n_i$ and $w_i$ represent respectively the momentum and winding numbers. At the localization locus the gauge fields are reduced to their holonomies on 
the worldsheet two-torus:
\begin{equation}
A^0=\frac{\pi\bar u}{i\tau_2}\, \text{d}w-\frac{\pi u}{i\tau_2}\, \text{d}\bar w\, .
\end{equation}
Plugging these expressions into~\eqref{eq:bosLag} leads to the zero modes part of the action:
\begin{equation}
\mathcal{S}_\text{bos}^0=\frac{\pi E_{ij}}{\tau_2}\Big\{(w_i\tau-n_i)(w_j\bar\tau-n_j)-v^i(w_j\bar\tau-n_j)u-v^j(w_i\bar\tau-n_i)u+v^i v^j u\bar u\Big\}\, .
\end{equation}
The partition function is given as a sum over the momenta and windings:
\begin{equation}
\mathcal{Z}_{\text{bos}}^0=\sum_{(w_i,n_i)\in\mathbb{Z}^{20}}\exp\left(-\mathcal{S}_\text{bos}^0\right)\, .
\end{equation}
Adopting obvious matrix notations, we can write the action as:
\begin{equation}
\mathcal{S}_\text{bos}^0=\frac{\pi}{\tau_2}\Big\{n\cdot Gn+F\cdot n+|\tau|^2w\cdot Gw-2\bar\tau u\, Gv\cdot w+u\bar u\, v\cdot Gv\Big\}\, ,
\end{equation}
where we have defined:
\begin{equation}
F:=-\big(\bar\tau E+\tau\bar E\big)w+2u\, Gv\, .
\end{equation}
After performing a Poisson resummation on each variable $n_i$, one gets:
\begin{multline}
\mathcal{Z}_{\text{bos}}^0=\frac{\sqrt{\tau_2}^{10}}{\sqrt{\text{det}\, G}}\, e^{-\frac{\pi}{\tau_2}v\cdot Gv\, u\bar u} \\ \sum_{(w,n)\in\mathbb{Z}^{20}}
 \exp \Big\{-\pi\tau_2\left(n-\frac{F}{2i\tau_2}\right)\cdot G^{-1}\left(n-\frac{F}{2i\tau_2}\right)
-\frac{\pi
}{\tau_2}\left(|\tau|^2w\cdot Gw-2\bar\tau u\, Gv\cdot w\right)\Big\}\, .
\end{multline}
Let us introduce the left and right momenta:
\begin{equation}
P_\textsc{l}=\frac{1}{\sqrt{2}}\, G^{-1}\big(n-(B-G)\, w\big) \ , \quad 
P_\textsc{r}=\frac{1}{\sqrt{2}}\, G^{-1}\big(n-(B+G)\, w\big)\, .
\end{equation}
One then has, after adding the contribution from the quantum fluctuations\footnote{We set $q:=\exp(2i\pi\tau)$.}:
\begin{equation}
\mathcal{Z}_{\text{bos}}=\frac{1}{|\eta(\tau)|^{20}}\, e^{-\frac{\pi}{\tau_2}v\cdot Gv\, (u\bar u-u^2)}\sum_{(w,n)\in\mathbb{Z}^{20}}
q^{\frac{1}{2}P_\textsc{l}^2}\bar q^{\frac{1}{2}P_\textsc{r}^2}e^{-2i\pi\sqrt{2}\, u\, v\cdot GP_\textsc{l}},
\end{equation}
with $P_\textsc{l}^2=P_\textsc{l}\cdot GP_\textsc{l}$ and $P_\textsc{r}^2=P_\textsc{r}\cdot GP_\textsc{r}$. Let us introduce the following $(20)\times (20)$ matrices:
\begin{equation}
\mathcal M=\begin{pmatrix}
G^{-1}&\ \, -G^{-1}B\\
BG^{-1}&\ \, G-BG^{-1}B
\end{pmatrix}\, ,\ \ \ \ 
\mathbb{I}=\begin{pmatrix}
0&\mathbb{I}_{10}\\
\mathbb{I}_{10}&0
\end{pmatrix}\, .
\end{equation}
In terms of these matrices, one has:
\begin{equation}
\frac{1}{2}P_\textsc{l}^2=\frac{1}{4}\begin{pmatrix}n&\, w\end{pmatrix}(\mathcal M+\mathbb{I})\begin{pmatrix}n\\w\end{pmatrix}\,  \ , \quad 
\frac{1}{2}P_\textsc{r}^2=\frac{1}{4}\begin{pmatrix}n&\, w\end{pmatrix}(\mathcal M-\mathbb{I})\begin{pmatrix}n\\w\end{pmatrix}\, .
\label{eq:quadraticMomenta}
\end{equation}

The spurious contributions to the path integral, a consequence of the formulation of the GLSM 
in terms of shift multiplets, are dealt with as follows. First, the real part of each complex boson $\omega^i$ 
gives a $V/(\sqrt{\tau_2}|\eta|^2)$ contribution, proportional to the infinite volume $V$ of their target space, 
which factorizes completely from the result. Second, the anti-holomorphic contribution of the right-moving part of $\text{Im}(b_n)$ is also completely factorized, given that its zero-modes contribution span an $E_8$ root lattice of signature $(0,8)$. 
Indeed by construction the 'Wilson lines' deformation do not involve this sub-lattice of the $(10,10)$ 
lattice corresponding to the 'extended' fiber, see eq.~(\ref{eq:extlattice}).

It leads  eventually to a expression similar to standard heterotic lattices with Wilson lines, 
in terms of $8$ complex  moduli $V^a$, together with an extra left coupling of the 
torus fiber to the worldsheet gauge holonomy:
\begin{align}
\mathcal{Z}_{\text{bos}}=&\frac{1}{\eta(\tau)^{18}\bar\eta(\bar\tau)^2}\, \exp\left\{-\frac{\pi}{\tau_2}\frac{U_2}{T_2}|\mathfrak{m}|^2\, (u\bar u-u^2)\right\}\times\nonumber\\
&\times\sum_{\substack{(n_1,n_2,w_1,w_2)\in\mathbb{Z}^{4},\\N\in\Gamma_{8,0}}}q^{\frac{1}{4}|p_\textsc{l}|^2}\bar q^{\frac{1}{4}|p_\textsc{r}|^2}
\exp\left(-2i\pi u\, \text{Re}\left(\mathfrak{m}\, \overline{p_\textsc{l}^0}\right)\right),
\end{align}
with the following standard complex expression for the momenta:
\begin{subequations}
\begin{align}
|p_\textsc{r}|^2&=\frac{1}{\left(T_2 U_2-\sum_a(V_2^a)^2\right)}\left|-n_1 T+n_2+w_1 U+w_2\left(TU-\sum_a(V^a)^2\right)+N_a V^a\right|^2\, ,\\
|p_\textsc{l}|^2&=|p_\textsc{r}|^2+4(n_1 w_1+n_2 w_2)+ N_a N^a\, ,
\end{align}
\end{subequations}
and where $p_\textsc{l}^0$ is the left-moving momentum along the two-torus, in the absence of Abelian bundle, written in complex notation:
\begin{equation}
\label{eq:p0def}
p_\textsc{l}^0:=\left.p_\textsc{l}\right|_{V^a=0}=\frac{1}{\sqrt{U_2T_2}}\big(-n_1 T+n_2+U(w_1+Tw_2)\big)\, .
\end{equation}

The relation between the complex Wilson line moduli $V^a=V_1^a+TV_2^a$ and the couplings $\beta_{\ell n}$ is then given by:
\begin{subequations}
\begin{align}
4iV_1^a&=\left(1+\frac{T_1^2}{T_2}\right)\beta_{1a}-\frac{T_1}{T_2}\beta_{2a}\, ,\\
4iV_2^a&=\frac{T_1}{T_2}\beta_{1a}-\frac{1}{T_2}\beta_{2a}\, .
\end{align}
\label{eq:vbeta}
\end{subequations}

Let us finally consider the contribution from the free fermions $\chi^i,\bar\chi^i$. 
After Wick rotation of~\cref{eq:ShiftLagrangian}, one has:
\begin{equation}
\mathcal{L}_\text{fer}=\frac{G_{ij}}{2}\, \bar\chi^i\, \partial\chi^j\, .
\end{equation}
On the other hand, the right-moving current is of the form:
\begin{equation}
\bar J=G_{ij}\, \bar\chi^i\chi^j+\ldots\, ,
\end{equation}
where the ellipsis stand for the contribution of all the other fields and possible $\mathcal{Q}$-exact terms. As discussed previously, 
a $(\bar J_0)^{10}$ allows to handle all the fermionic zero-modes originating from the torus fiber and Wilson lines 
fermions, see~\cref{eq:FYpart}, and one obtains a $\bar\eta(\bar\tau)^2$ contribution for each of the $10$ free fermions which 
is canceled by the  $1/\bar\eta^{20}$ in the definition of the intermediate supersymmetric index defined in~(\ref{eq:FYpartint}). 

\subsubsection{The result}

Assembling all pieces together, namely the contributions of the chiral and Fermi multiplets from the $K3$ base, of the $U(1)$ vector multiplets, 
and those from the torus fiber and 'Wilson lines', one obtains:
\begin{align}
Z_\textsc{fy}^\textsc{w} (\tau,\bar \tau, z) =& \pm
\left(-2i\pi\eta(\tau)^{2}\right) \times \nonumber \\
&\sum_{u^{\star}\in \mathcal{M}_{\text{sing}}^{\pm}}\oint_{u=u^{\star}}\text{d} u\, 
\Bigg\{
\prod_{\Phi_{i}}\frac{i\eta(\tau)}{\vartheta_{1}(\tau|Q_{i}u +q_{i}^{\textsc{l}}z)}
\prod_{\Gamma_{a}}\frac{i\vartheta_{1}(\tau|Q_{a} u +q^{\textsc{l}}_{a}z)}{\eta(\tau)}\nonumber\\
&\sum_{(p_\textsc{l},p_\textsc{r})\in\Gamma_{10,2}}\frac{q^{\frac{1}{4}|p_\textsc{l}|^2}}{\eta(\tau)^{18}}
\frac{\bar q^{\frac{1}{4}|p_\textsc{r}|^2}}{\bar\eta(\bar\tau)^2}\exp\left(-2i\pi u\, \text{Re}\left(\mathfrak{m}\, \overline{p_\textsc{l}^0}\right)\right)\Bigg\}\, ,
\label{eq:oneloopdetrank}
\end{align}
Thanks to the tadpole condition~((\ref{eq:tad})), the global factor from the determinants
\begin{equation*}
\exp\left\{-\frac{\pi\left(u^2-u\bar u\right)}{\tau_2}\left(\sum_{\text{chiral}}Q^2-\sum_{\text{fermi}}Q^2-\frac{2U_2}{T_2}|\mathfrak{m}|^2\right)\right\}\, ,
\end{equation*}
which is non holomorphic in the gauge field holonomy (hence potentially forbidding the reduction to a contour integral) vanishes. There are also similar non-holomorphic 
terms involving the $U(1)_\textsc{l}$ global charges which vanish owing to the cancellation of the corresponding (mixed) anomalies. 

A consistent choice of global charges, as we have already discussed in~\cite{Israel:2015aea}, is to assign $U(1)_\textsc{l}$ charge $+1$ to the chiral multiplet $P$, charges 
$-1$ to the Fermi multiplets $\Gamma_a$, both appearing in the superpotential term~(\ref{eq:pterm}), and vanishing $U(1)_\textsc{l}$ charge to all other multiplets.

In the formula~(\ref{eq:oneloopdetrank}), $\mathcal{M}_{\text{sing}}^{\pm}$ corresponds to one of two sets of singularities in the $u$ plane 
for the determinants of chiral multiplets (from the $K3$ base), of positive or negative gauge charge respectively~\cite{2015CMaPh.333.1241B}. 
Both choices are equivalent since the sum of residues of a meromorphic function on the torus vanishes, however the 
natural interpretation of the formula is different in both cases. In general, the expression obtained from $\mathcal{M}_{\text{sing}}^{+}$ would correspond to a Landau--Ginzburg type of computation. 

Picking up $\mathcal{M}_{\text{sing}}^{-}$, giving typically a contour integral around a pole at the origin, and 
provides the result that one would obtain by a direct computation in the geometrical 'phase',  
flowing in the IR to a (large volume) non-linear sigma-model. In the next section, we will provide a corresponding 
geometrical formula for the index, while the equivalence between both expressions is proven in appendix~\ref{app:ProofGeom}, when the $K3$ surface is a subvariety of a weighted projective space.

In the computation of the index that we have presented in this section, we considered a $U(1)$ worldsheet gauge group 
for clarity. The result can be generalized for higher rank gauge groups in terms of a 
sum of Jeffrey-Kirwan residues using the results from \cite{2015CMaPh.333.1241B}, as we have done in~\cite{Israel:2015aea}. Instead of going along this route we will instead move to the geometrical formulation 
of the supersymmetric index, which is expected to be equally valid for any formulation, or UV completion, of the worldsheet theory underlying the 
$\mathcal N=2$ compactifications with torsion.


\section{Geometrical formulation of the dressed elliptic genus}
\label{sec:GeomForm}

In this section we provide a geometrical formula for the dressed elliptic genus, whose 
Hamiltonian definition is given by eq.~(\ref{eq:FYpart}), associated with a heterotic 
compactification given by the principal bundle $T^2\hookrightarrow X\stackrel{\pi}{\to}\mathcal{S}$, together with a 
 gauge bundle compatible with $\mathcal{N}=2$ space-time supersymmetry.

We summarize here the 
relevant bundle data of such non-K\"ahler heterotic compactification:
\begin{itemize}
\item The holomorphic tangent bundle $T_\mathcal{S}$ over the base, with $c_1 (T_\mathcal{S}) = 0$, 
\item A rank $r$ stable holomorphic vector bundle $\mathcal V$ over $\mathcal{S}$, with $c_1 (\mathcal{V}) = 0$, 
whose pullback provides a  gauge bundle on $X$ compatible with supersymmetry, 
\item A pair of anti-self-dual two-forms $\omega_1$ and $\omega_2$ on $\mathcal{S}$, defining 
two equivalence classes in $H^{2}(\mathcal{S},\mathbb{Z})\cap 
H^{1,1}_{\bar\partial}(\mathcal{S})$,
\item A heterotic Narain lattice $\Gamma (T,U,V)$ of signature $(10,2)$, with $T$ and $U$ belonging to the same imaginary quadratic number field $\mathbb{Q}(\sqrt{D})$.
\end{itemize}

In appendix~\ref{app:rat}, we discuss in more detail the quantization of the torus moduli, and the compatibility between the choice of 
rational Narain lattice and of the pair of two-forms $(\omega_1,\omega_2)$. These anti-self-dual two-forms 
generate a rank two\footnote{We don't consider the degenerate case where $\omega_1$ and $\omega_2$ are colinear.} 
even negative-definite lattice $\Gamma_\omega$, which is a sub-lattice of $H^2 (\mathcal{S},\mathbb{Z})$.

The integrated Bianchi identity~(\ref{eq:tad}), or tadpole condition, can be written in the following compact way: 
\begin{equation}
\label{eq:tad2}
N-24 = -\int_{\mathcal{S}}p_{\omega}\wedge\star_{\scriptscriptstyle \mathcal{S}}\, \bar{p}_{\omega} \, ,
\end{equation}
where $N$ is the instanton number of the gauge bundle, defined as 
\begin{equation}
\label{eq:instantonNumber}
N = -\int_\mathcal{S} \text{ch}_2 (\mathcal{V})\, .
\end{equation}
In the present context, $N$ is an integer between 0 and 24, the latter case corresponding to  a $K3\times T^2$ compactification. 

We have also introduced in eq.~(\ref{eq:tad2}) the following two-dimensional vector of two-forms, 
built by embedding $(\omega_1,\omega_2)$ into 
the lattice of the two-torus fiber, given in complex notation as:
\begin{equation}
\label{eq:p_omega_def}
p_\omega:=\sqrt{\frac{U_2}{T_2}}\, (\omega_1+T\omega_2)\, .
\end{equation}
This vector as it is belongs to a formal extension, over $H^2(\mathcal{S},\mathbb{Z})$, 
of the winding sub-lattice of the $\Gamma^{2,2} (T,U)$ toroidal lattice. As explained 
in appendix~\ref{app:rat}, the compatibility conditions~(\ref{eq:quantcondforms}) 
between $\omega_{1,2}$ and the lattice ensure that it actually belongs to (a formal extension of) 
the left lattice $\Gamma_\textsc{l}$ of the two-torus. Notice that $p_\omega$ 
involves the moduli $U$ and $T$ of the torus with the 'Wilson lines' turned off, and not 
those corresponding to the physical Kaluza-Klein metric on $T^2$. 

As for the ordinary elliptic genera of holomorphic vector bundles over Calabi-Yau manifolds~\cite{Witten:1986bf}, we define first 
the formal power series with bundle coefficients
\begin{equation}\label{eq:formalbundle}
\mathbb{E}_{q,w} = \bigotimes_{n=0}^\infty \bigwedge\nolimits_{-w q^n} \mathcal V^\star \otimes  
\bigotimes_{n=1}^\infty \bigwedge\nolimits_{-w^{-1} q^n} \mathcal  V \otimes 
\bigotimes_{n=1}^\infty S_{q^n} T^\star_\mathcal{S} \otimes 
\bigotimes_{n=1}^\infty S_{q^n} T_\mathcal{S}\, ,
\end{equation}
where 
\begin{equation}
\bigwedge\nolimits_{t} \mathcal V = \sum_{k=0}^{\infty}\, t^k\, \bigwedge\nolimits^k \mathcal V\ , \quad 
S_t \mathcal T_\mathcal{S} =\sum_{k=0}^{\infty}\, t^k \, S^k \, \mathcal T_\mathcal{S}\, ,
\end{equation}
$\bigwedge\nolimits^k$ and $S^k$ being respectively  the $k$-th exterior product and the $k$-th symmetric product.

\subsection{Modified Euler characteristic}

Using the notations introduced above, we define the dressed elliptic genus of a  holomorphic vector bundle 
$\mathcal{V}$ over a $K3$ surface $\mathcal S$, 
with a given $(10,2)$ lattice comprising the $T^2$ fiber of the principal bundle $T^2\hookrightarrow X\stackrel{\pi}{\to}\mathcal{S}$, 
and the set of Abelian connections over $X$, as the following modified Euler characteristic:
\begin{equation}
\label{eq:geom}
Z_{\textsc{fy}}^\textsc{w} \left(\left.X,\mathcal{V},\omega\right|\tau,\bar{\tau},z\right)=
q^{\frac{r-2}{12}}y^{\frac{r}{2}}\int_{\mathcal{S}}\text{ch}\left(\mathbb{E}_{q,y}\right)
\text{td}\left(T_\mathcal{S}\right)\sum_{(p_\textsc{l},p_\textsc{r})\in\Gamma_{10,2}}
\frac{q^{\frac{1}{4}|p_\textsc{l}|^2}}{\eta(\tau)^{18}}
\frac{\bar q^{\frac{1}{4}|p_\textsc{r}|^2}}{\bar\eta(\bar\tau)^2}e^{-\, \text{Re}\left(p_\omega\, \overline{p_\textsc{l}^0}\right)}
\, ,
\end{equation}
where $\overline{p_\textsc{l}^0}$ is defined in eq.~(\ref{eq:p0def}).

The proof that this formula actually coincides with the GLSM result, eq.~(\ref{eq:oneloopdetrank}), 
is given in appendix~\ref{app:ProofGeom}, in the cases where $\mathcal{S}$ 
is constructed as a subvariety of a weighted projective space $V=\mathbb{P}^n(q_0,...,q_n)$.

Let $c(\mathcal{T}_\mathcal{S})=\prod_{i=1}^{2}(1+\nu_{i})$ 
and $c(\mathcal V)=\prod_{a=1}^{r}(1+\xi_{a})$ denote the total Chern classes of the respective bundles, 
making use of the splitting principle. One can write
\begin{equation}
Z_\textsc{fy}^\textsc{w} (X,\mathcal V,\omega|\tau,\bar \tau,z) =\int_{\mathcal{S}}G(\tau,\bar{\tau},z,\nu,\xi,p_\omega)\, 
,\label{eq:geometric_formula2}
\end{equation}
with integrand
\begin{align}
\label{eq:integrand}
G(\tau,\bar{\tau},z,\nu,\xi,p_\omega)=&\prod_{a=1}^{r}\frac{i\theta_{1}\left(\tau\left|\frac{\xi_{a}}{2i\pi}-z\right.\right)}{\eta(\tau)}
\prod_{i=1}^{2}\frac{\eta(\tau)\nu_{i}}{i\theta_{1}(\tau\left|\frac{\nu_{i}}{2i\pi}\right.)}\times\nonumber\\
&\times\sum_{(p_\textsc{l},p_\textsc{r})\in\Gamma_{10,2}}\frac{q^{\frac{1}{4}|p_\textsc{l}|^2}}{\eta(\tau)^{18}}\frac{\bar 
q^{\frac{1}{4}|p_\textsc{r}|^2}}{\bar\eta(\bar\tau)^2}e^{-\, \text{Re}\left(p_\omega\, \overline{p_\textsc{l}^0}\right)}\, .
\end{align}

\subsection{Modular properties}

The ordinary elliptic genus  of a rank $r$ holomorphic vector bundle of vanishing first Chern class over a 
K3 surface $\mathcal S$, satisfying the condition $c_2 (\mathcal V) = c_2 (T_\mathcal{S})$, 
is a weak Jacobi form of weight 0 and index $r/2$ with the same character, or multiplier system, as $(\theta_1/\eta)^{r-2}$.

Recall that a weak Jacobi form of weight $k$, index $m$ and character $\chi$ is 
a holomorphic map $\phi:\mathbb{H}\times\mathbb{C}\rightarrow\mathbb{C}$ such that 
$$\forall\gamma=\begin{psmallmatrix}a&b\\c&d\end{psmallmatrix} \in SL_2(\mathbb{Z})\, ,\ 
\phi\left(\frac{a\tau+b}{c\tau+d},\frac{z}{c\tau+d}\right)=\chi(\gamma)(c\tau+d)^k 
e^{2i\pi m\frac{cz^2}{c\tau+d}}\phi(\tau,z)$$ and such that  
$$\forall\lambda,\mu\in\mathbb{Z}\, ,\ 
\phi(\tau,z+\lambda\tau+\mu)=(-1)^{2t(\lambda+\mu)}e^{-2i\pi m(\lambda^2\tau+2\lambda z)}\phi(\tau,z)\, .$$ Furthermore its 
Fourier expansion should be of the form 
$$\phi(\tau,z)=\sum_{\substack{n\geqslant 0\\l\in\mathbb{Z}+t}}a_{n,l}q^n y^l\, .$$

The dressed elliptic genus $Z_\textsc{fy}^\textsc{w}$ that we have 
defined for non-K\"ahler  $T^2\hookrightarrow X\stackrel{\pi}{\to}\mathcal{S}$ principal bundles, 
though non-holomorphic in $\tau$ by construction, is holomorphic in the $z$ variable and transforms also as a weak 
Jacobi form\footnote{If the fiber of the principal bundle 
is one-dimensional ($i.e.$ an $S^1$ rather than a $T^2$), in the absence of 'Wilson lines', the non-trivial part of the index 
$\int_{\mathcal{S}}
\prod_{a=1}^{r}\frac{i\theta_{1}\left(\tau\left|\frac{\xi_{a}}{2i\pi}-z\right.\right)}{\eta(\tau)}
\prod_{i=1}^{2}\frac{\eta(\tau)\nu_{i}}{i\theta_{1}(\tau\left|\frac{\nu_{i}}{2i\pi}\right.)}
\sum_{\mu \in \mathbb{Z}_{2k}} \Theta_{\mu,k} \left(\tau | \tfrac{m^1\omega_1 }{2i\pi} \right) \Theta_{\varphi(\mu),k} (-\bar \tau | 0)$ 
has some similarity with skew-holomorphic Jacobi forms, defined by Skoruppa~\cite{Skoruppa}, but fails to satisfy a heat equation.}
of weight $0$ and index $\frac{r}{2}$, with the same character as $(\theta_1/\eta)^{r-2} E_4/\eta^8$, as can be seen by a trivial 
generalization of the computation that we have presented in~\cite{Israel:2015aea}. 

\subsection{Decomposition into weak Jacobi forms}

An explicit expression of the dressed elliptic genus can then be obtained, with minimal 
knowledge of the underlying geometrical data. The following formula holds~\cite{Gritsenko:1999nm}:
\begin{equation}
\theta_1(\tau|z+\xi)=\exp\left\{-\frac{\pi^2}{6}E_2(\tau)\xi^2+\frac{\theta_1'(\tau|z)}{\theta_1(\tau|z)}\xi-\sum_{n\geqslant 2}\wp^{(n-2)}(\tau,z)\frac{\xi^n}{n!}\right\}\theta_1(\tau|z)\, ,
\end{equation}
where  $\wp$ is the Weierstrass elliptic function and $\wp^{(n)}:=\frac{\partial^n}{\partial z^n}\wp$.
Expanding the integrand~(\ref{eq:integrand}) and keeping only the top degree form terms, one obtains
\begin{multline}
\label{eq:geometric_formula3}
Z_\textsc{fy}^\textsc{w}=
(-i)^r\frac{\theta_1(\tau|z)^r}{\eta(\tau)^{r+4}}\sum_{(p_\textsc{l},p_\textsc{r})\in\Gamma_{10,2}}\frac{q^{\frac{1}{4}|p_\textsc{l}|^2}}{\eta(\tau)^{18}}\frac{\bar q^{\frac{1}{4}|p_\textsc{r}|^2}}{\bar\eta(\bar\tau)^2}\times\\
\times\int_\mathcal{S}\left\{-\frac{E_2(\tau)}{24}\sum_i \nu_i^2+\left(\frac{E_2(\tau)}{24}-\frac{\wp(\tau,z)}{2(2i\pi)^2}\right)\sum_a \xi_a^2+\frac{1}{2}\text{Re}\left(p_\omega\, \overline{p_\textsc{l}^0}\right)^2\right\}\, ,
\end{multline}
where we have used the fact that the first Chern class of the holomorphic vector bundle $\text{c}_{1}(\mathcal{V})$ vanishes. Using the definition of the instanton number~\eqref{eq:instantonNumber}, 
the fact that $\int_\mathcal{S}\text{ch}_2 (\mathcal{T_\mathcal{S}})=-24$, and that the ordinary elliptic genus of a $(4,4)$ non-linear  
sigma-model on $K3$, namely
\begin{equation}
Z_{\text{ell}}^{\scriptscriptstyle K3}(\tau,z)=8\left\{\left(\frac{\theta_2(\tau|z)}{\theta_2(\tau|0)}\right)^2
+\left(\frac{\theta_3(\tau|z)}{\theta_3(\tau|0)}\right)^2+\left(\frac{\theta_4(\tau|z)}{\theta_4(\tau|0)}\right)^2\right\}\, ,
\end{equation}
is related to the Weierstrass $\wp$-function by the following formula:
\begin{equation}
Z_{\text{ell}}^{\scriptscriptstyle K3}(\tau,z)=\frac{6}{\pi^2}\frac{\wp(\tau,z)\theta_1(\tau|z)^2}{\eta(\tau)^6}\, ,
\end{equation}
one can write the index as sum of three terms in the following way:
\begin{multline}
\label{eq:Index3terms}
Z_\textsc{fy}^\textsc{w}=
-(-i)^r\sum_{(p_\textsc{l},p_\textsc{r})\in\Gamma_{10,2}}\frac{q^{\frac{1}{4}|p_\textsc{l}|^2}}{\eta(\tau)^{18}}\frac{\bar q^{\frac{1}{4}|p_\textsc{r}|^2}}{\bar\eta(\bar\tau)^2}\times\\
\left(\frac{\vartheta_1(\tau|z)}{\eta(\tau)}\right)^{r-2}
\left\{\frac{N}{24}
Z_{\text{ell}}^{\scriptscriptstyle K3}(\tau,z)
+\frac{N-24}{12}\frac{\theta_1(\tau|z)^2}{\eta(\tau)^{6}}\, E_2(\tau)
-\frac{\theta_1(\tau|z)^2}{2\, \eta(\tau)^{6}}\int_{\mathcal{S}}\text{Re}\left(p_\omega\, \overline{p_\textsc{l}^0}\right)^2\right\}\, .
\end{multline}
This expression can be rewritten as:
\begin{equation}
\label{eq:geometric_formula4}
Z_\textsc{fy}^\textsc{w}=
-(-i)^r\sum_{(p_\textsc{l},p_\textsc{r})\in\Gamma_{10,2}}\frac{q^{\frac{1}{4}|p_\textsc{l}|^2}}{\eta(\tau)^{18}}\frac{\bar q^{\frac{1}{4}|p_\textsc{r}|^2}}{\bar\eta(\bar\tau)^2}
\left(\frac{\vartheta_1(\tau|z)}{\eta(\tau)}\right)^{r-2}\mathcal{Z}_N(\tau,z,p_\textsc{l})\, ,
\end{equation}
where:
\begin{equation}
\label{eq:toBeExpanded}
\mathcal{Z}_N(\tau,z,p_\textsc{l})=\frac{N}{12}\, \phi_{0,1}(\tau,z)+\left(-\frac{N-24}{12}\, E_2(\tau)
+m(p_\textsc{l},\omega)\right)\phi_{-2,1}(\tau,z)
\end{equation}
with the standard weak Jacobi forms of index 1:
\begin{align}
\phi_{0,1}(\tau,z)&=4\left(\frac{\theta_2(\tau|z)^2}{\theta_2(\tau|0)^2}+\frac{\theta_3(\tau|z)^2}{\theta_3(\tau|0)^2}+\frac{\theta_4(\tau|z)^2}{\theta_4(\tau|0)^2}\right)\, ,\nonumber\\
\phi_{-2,1}(\tau,z)&=-\frac{\theta_1(\tau|z)^2}{\eta(\tau)^6}\, ,
\end{align}
and where we have defined:\footnote{$m(p_\textsc{l},\omega)$ is an integer since the Picard lattice $\text{Pic}(\mathcal{S})=H^{2}(\mathcal{S},\mathbb{Z})\cap 
H^{1,1}_{\bar\partial}(\mathcal{S})$ is even.}
\begin{equation}
\label{eq:mdef}
m(p_\textsc{l},\omega):=\frac{1}{2}\int_{\mathcal{S}}\text{Re}\left(p_\omega\, \overline{p_\textsc{l}^0}\right)^2\, .
\end{equation}

The expression~(\ref{eq:toBeExpanded}) that we have obtained shows that the index depends  on the vector bundle 
$\mathcal V$ only through its instanton number $N$. The data characterizing the principal two-torus bundle is encoded in~(\ref{eq:mdef}), that intertwines $\Gamma_\omega$ with $\Gamma_{2,2} (T,U)$. Equation~(\ref{eq:toBeExpanded}) is a good starting point to compute the threshold 
corrections to the gauge and gravitational couplings, that will be given in~\cite{AIS}.


\section{Decomposition into \texorpdfstring{$\mathcal{N}=4$}{n=4} characters and moonshine properties of the index (or the lack of it)}
\label{sec:ExpansionIndex}

Non-linear sigma-models on $K3$ with $(4,4)$ supersymmetry have a rather mysterious relationship with the Mathieu group 
$\mathbb{M}_{24}$, that was first observed in~\cite{Eguchi:2010ej} (and explored later on by many authors), 
by expanding the elliptic genus of the former into characters of the $\mathcal{N}=4$ superconformal algebra:
\begin{equation}
Z_{\text{ell}}^{\scriptscriptstyle K3}(\tau,z)=24\, \text{ch}_{h=1/4,l=0}(\tau,z)+\sum_{n=0}^{\infty} A_n\, \text{ch}_{h=k+1/4,l=1/2}(\tau,z)\, ,
\end{equation}
where $\text{ch}_{h=1/4,l=0}$ is the character of the massless representation of isospin zero and 
$\text{ch}_{h=k+1/4,l=1/2}$ are characters of massive representations of isospin one-half. The coefficients $\{ A_n \}$ of the expansion are indeed related to dimensions of $\mathbb{M}_{24}$ irreducible 
representations as $A_0 = - 2$, $A_1 = 90 = 45 + \overline{45}$, etc...

This Mathieu moonshine can be extended to $K3$ compactifications with arbitrary gauge 
bundles~\cite{2013JHEP...09..030C}, despite the fact that the underlying 
two-dimensional theory has only $(0,4)$ supersymmetry (hence no $\mathcal{N}=4$ on the holomorphic side) 
thanks to the universality properties of the new supersymmetric index dictated by its modular properties~\cite{Kiritsis:1996dn}. 
It is therefore legitimate to investigate possible moonshine phenomena for the torsional 
compactifications investigated here. A first step is naturally to look for possible hints of relations with the group $\mathbb{M}_{24}$. 

One can actually expand $\mathcal{Z}_N(\tau,z,p_\textsc{l})$, the summand appearing in 
the dressed elliptic genus~(\ref{eq:geometric_formula4}), 
in terms of $\mathcal{N}=4$ characters as follows:
\begin{equation}
\label{eq:expn4}
\mathcal{Z}_N(\tau,z,p_\textsc{l})=N\, \text{ch}_{h=1/4,l=0}(\tau,z)+\sum_{n=0}^{\infty} \tilde{A}_n\, \text{ch}_{h=n+1/4,l=1/2}(\tau,z)\, ,
\end{equation}
with 
\begin{equation}
\tilde{A}_n(N,p_\textsc{l},\omega)=\frac{N}{24}A_n+\frac{N-24}{12}\, B_n-m(p_\textsc{l},\omega)\, C_n\, ,
\end{equation}
where the integer coefficients $B_n$ and $C_n$ are defined by the expansions:
\begin{subequations}
\begin{align}
\frac{q^{1/8}E_2}{\eta^3}&=\sum_{k=0}^{\infty}B_k\, q^k\, ,\\
\frac{q^{1/8}}{\eta^3}&=\sum_{k=0}^{\infty}C_k\, q^k\, .
\label{eq:cndef}
\end{align}
\end{subequations}

In particular the $\{ C_n\}$ give the number of partition into three kinds of integers. They 
can be expressed by the following recursive relations:
\begin{align}
&C_0=1\, ,\nonumber\\
&C_n=\frac{3}{k}\sum_{l=0}^{n-1}\, C_\ell\, \sigma_1(n-l),\ \forall n\in\mathbb{N}^* ,
\end{align}
where $\sigma_1(n)=\sum_{d|n}d$.
Moreover, one can expand the Eisenstein series $E_2$ as
\begin{equation}
E_2(\tau)=1-24\sum_{k=1}^{\infty}\sigma_1(k)\, q^k\, .
\end{equation}
One thus has the following relation between these two sequences of coefficients:
\begin{align}
\forall n\in\mathbb{N},\ B_n&=\, C_n-24\sum_{l=0}^{n-1}C_\ell\, \sigma_1(n-l)\nonumber\\
&=\, (1-8n)\, C_n\, .
\end{align}

Using these relations, one obtains the final expression for the coefficients $\{\tilde{A}_n\}$ of the 
expansion into $N=4$ characters, eq.~(\ref{eq:expn4}), as 
\begin{equation}
\label{eq:coeffn4}
\tilde{A}_n = 2 (8n-1) C_n + N\, \frac{A_n -2 (8n-1)C_n}{24} - m(p_\textsc{l},\omega) C_n \, . 
\end{equation}
The first coefficients of this expansion are explicitly:
\begin{equation}
{\setlength\arraycolsep{10pt}
\begin{array}{|c||c|}
\hline
n&\tilde{A}_n \vphantom{A^{A^{A^A}}}\\
\hline\hline
0&-2-m(p_\textsc{l})\\
\hline
1&42+2N-3\, m(p_\textsc{l})\\
\hline
2&270+8N-9\, m(p_\textsc{l})\\
\hline
3&1012 + 22 N-22\, m(p_\textsc{l})\\
\hline
4&3162 + 58 N-51\, m(p_\textsc{l})\\
\hline
5&8424 + 132 N-108\, m(p_\textsc{l})\\
\hline
6 &20774 + 294 N-221\, m(p_\textsc{l})\\
\hline
7 &47190 + 604 N-429\, m(p_\textsc{l})\\
\hline
8 &102060 + 1210 N-810\, m(p_\textsc{l})\\
\hline
9 &210018 + 2318 N-1479\, m(p_\textsc{l})\\
\hline
10 &417120 + 4334 N -2640\, m(p_\textsc{l})\\
\hline
\end{array}
}
\end{equation}

All the coefficients $\{\tilde{A}_n \}$ are integer numbers, which is not obvious from their expression~(\ref{eq:coeffn4}). This 
follows from a quite intriguing property, which may shed some light on the ordinary Mathieu moonshine, namely that:
\begin{equation}
\label{eq:congFaible}
\forall n \in \mathbb{N} \ , \quad A_n -2 (8n-1)C_n \equiv 0 \mod 24 \, ,
\end{equation}
where the $\{ A_n \}$ are the coefficients of the expansion of the $K3$ elliptic genus into $\mathcal{N}=4$ representations --~hence 
encode the information about $\mathbb{M}_{24}$ representations~-- and where $C_n$ are defined by~(\ref{eq:cndef}).
In fact one has a stronger result:
\begin{equation}
\label{eq:congForte}
\forall n \in \mathbb{N} \ , \quad A_n -2 (8n-1)C_n \equiv 0 \mod 48 \, .
\end{equation}
To see it, let us define:
\begin{equation}
A(q):=\sum_{n=0}^{\infty}A_nq^n\, ,\ \ \ \ \ C(q):=\frac{q^{1/8}}{\eta(q)^3}\, ,
\end{equation}
and consider the function:
\begin{equation}
\nu(q):=A(q)-2\left(8q\frac{\partial}{\partial q}-1\right)C(q)\, .
\end{equation}
This function then has the following $q$-expansion:
\begin{equation}
\nu(q)=\sum_{n=0}^\infty \Big(A_n -2 (8n-1)C_n\Big)q^n\, .
\end{equation}
Using the fact that $q\partial_q=(2i\pi)^{-1}\partial_\tau$ and that:
\begin{equation}
E_2(\tau)=\frac{12}{i\pi}\frac{\partial}{\partial\tau}\eta(\tau)\, ,
\end{equation}
one computes
\begin{equation}
\nu(q)=A(q)+2\, \frac{q^{1/8}}{\eta(q)^3}\, E_2(q)\, .
\end{equation}
One can then use eq. (7.16) of \cite{Dabholkar:2012nd} to obtain:
\begin{equation}
\nu(q)=48\, \frac{q^{1/8}}{\eta(q)^3}\, F^{(2)}(q)\, ,
\end{equation}
with:
\begin{equation}
F^{(2)}(q)=\sum_{\substack{0<m<n\\n\not\equiv m\text{ mod 2}}}(-1)^n\, m\, q^{\frac{mn}{2}}\, ,
\end{equation}
the relevant point being that $48$ divides $\nu(q)$. It may be possible to
 generate other identities of the type \cref{eq:congForte} by considering 
twists by insertion of $\mathbb{M}_{24}$ elements.

Coming back to the possible moonshine behavior of the index~(\ref{eq:expn4}), given that the coefficients $m (p_\textsc{l})$ can be arbitrary large (negative) integers, 
depending on the left-moving momentum along the two-torus fiber, the decomposition of the coefficients 
$\{ \tilde{A}_n \}$ into dimensions of irreducible representations of $\mathbb{M}_{24}$, or any other sporadic group, 
is far from obvious. If these coefficients were corresponding each to the dimension of a 
given representation, the term in $m(p_\textsc{l},\omega)$ could indicate the number of times such representation appears in the module for this $p_\textsc{l}$.

By considering the problem from another point of view, one could notice that the quantity $\mathcal{Z}_N(\tau,z,p_\textsc{l})$ appearing in~\cref{eq:toBeExpanded} is actually 
similar to a twining partition function in the context of the standard Mathieu moonshine \cite{Gaberdiel:2012gf,Gannon:2012ck}, if one sets aside the contribution 
in $m(p_\textsc{l},\omega)$ which involves the torus fiber. Further investigations are under way~\cite{HIPS}.

\section{Conclusion}

In this article we have completed the computation of the new supersymmetric index of $\mathcal{N}=2$ heterotic non-K\"ahler compactifications with torsion initiated in~\cite{Israel:2015aea}, by considering a larger class gauge bundles compatible with supersymmetry, tensoring the pullback of a holomorphic stable vector bundle over the $K3$ base with flat Abelian connections over the total space $X$ of the 
principal bundle $T^2\hookrightarrow X\rightarrow K3$, that would reduce to ordinary Wilson lines along the torus for 
factorized $K3\times T^2$ compactifications.  

We have obtained this result starting with a $(0,2)$ GLSM with torsion of the sort introduced in~\cite{Adams:2006kb}, suitably modified in order to accommodate for the Abelian bundles that 
were not considered in the original construction, and using supersymmetric localization. The index is naturally expressed in terms of a non-holomorphic dressing of the anomalous $K3$ elliptic 
genus by the contribution of the torus fiber.  

We then provided a mathematical definition of this dressed elliptic genus in terms of bundle data, which is completely generic. Starting from this geometrical formulation we 
exhibited an expansion in terms of $\mathcal N =4$ superconformal characters whose coefficients were shown to be integers, by proving a congruence identity relating the dimensions of $\mathbb{M}_{24}$ 
representations appearing in the moonshine module of the $K3$ elliptic genus  to a combinatoric factor describing the tri-partition of integers.

As mentioned in the introduction, the new supersymmetric index is the natural starting point for computing the gauge and gravitational one-loop threshold corrections appearing in 
the low-energy four-dimensional effective action of $\mathcal{N}=2$ heterotic compactifications. Given that the torsional compactifications considered in the present work represent a large fraction of those, including as 
a subset the familiar $K3\times T^2$  compactifications, computing these thresholds is a rather important task. 

Starting from~\cref{eq:Index3terms} in the present work, one can reach an expression in terms of standard weak almost holomorphic modular forms for the 
threshold corrections, and exploit the whole machinery developped in~\cite{Angelantonj:2011br,Angelantonj:2012gw} for performing modular integrals by 
unfolding the integration domain against Niebur-Poincar\'e series. These results will be presented in a forthcoming publication~\cite{AIS}.

The computation of the heterotic threshold corrections will also shed light on $\mathcal{N}=2$ type IIA/heterotic dualities. Potential duals of 
torsional heterotic compactifications were proposed in~\cite{Melnikov:2012cv}, as Calabi-Yau three-folds admitting a K3 fibration without compatible 
elliptic fibration with section. Given that the Abelian bundle moduli are not quantized by $H$-flux, unlike the $T$ and $U$ moduli of the two-torus fiber, 
one can in principle compute the associated prepotential governing the complex structure moduli from the heterotic threshold corrections, and compare with the type II 
expectations, as was done for $K3\times T^2$ in~\cite{2013JHEP...09..030C}. We plan to perform such quantitative checks, that would extend 
significantly our current knowledge of type IIA/heterotic $\mathcal{N}=2$ dualities, in the near future.

\section*{Acknowledgments}

We thank Carlo Angelantonj, Miranda Cheng, Sarah Harrison, Chris Hull, Shamit Kachru, Ilarion Melnikov, Ruben 
Minasian, Natalie Paquette and Ronen Plesser for discussions and correspondence. This work was conducted within the 
ILP LABEX (ANR-10-LABX-63) supported by French state funds managed by the 
ANR (ANR-11-IDEX-0004-02) and by the project QHNS in the program ANR Blanc 
SIMI5 of Agence National de la Recherche. 
 

\appendix

\section{(0,2) superspace}
\label{app:Superspace}

Minkowskian $(0,2)$ superspace is spanned by the coordinates $(\sigma^+,\sigma^-, \theta,\bar \theta)$. We define the superspace derivatives and super-charges as follows:
\begin{subequations}
\begin{align}
\mathcal{Q}_+ &= \partial_{\theta} +i \bar\theta \partial_+ \quad , \qquad \bar{\mathcal{Q}}_+ = -\partial_{\bar\theta} -i \theta  \partial_+ \, , \\
D_+ &= \partial_{\theta} -i \bar\theta \partial_+ \quad , \qquad \bar{D}_+ = -\partial_{\bar\theta} + i\theta \partial_+  \, .
\end{align}
\end{subequations}
The non-trivial anti-commutators are then
\begin{equation}
\{ \bar{D}_+,D_+\} = 2 i \partial_+  \quad , \qquad \{ \bar{\mathcal{Q}}_+,\mathcal{Q}_+\} = -2i \partial_+
\end{equation}

\paragraph{Chiral superfields} are defined by the constraint
\begin{equation}
\bar{D}_+ \Phi = 0 \ \implies \ 
\Phi = \phi + \sqrt{2}\theta \lambda_+ -i \theta \bar\theta  \partial_+ \phi
\end{equation}

\paragraph{Fermi superfields} have as a bottom component a left-moving fermion. They satisfy generically the constraint
\begin{equation}
\label{fermiconstr}
\bar{D}_+ \Gamma = \sqrt{2} E (\Phi_i) \, ,
\end{equation}
where $E$ is an holomorphic function. Hence they have the component expansion
\begin{equation}
\Gamma =\gamma_- + \sqrt{2} \theta G - \sqrt{2} \bar\theta E(\Phi_i) -i \theta \bar\theta \partial_+ \gamma_-\, ,
\end{equation}
where $G$ is an auxiliary field.

\paragraph{Gauge multiplets} are actually described by a pair of $(0,2)$ superfields, namely $\mathcal{A}$ and 
$\mathcal{V} $. 
Super-gauge transformations act as
\begin{equation}
\mathcal{A} \to \mathcal{A} +\frac{i}{2}(\bar \Xi - \Xi) \quad , \qquad 
\mathcal{V}  \to \mathcal{V}  -\frac{1}{2} \partial_- (\Xi + \bar \Xi) 
\end{equation}
where $\Xi$ is a chiral superfield. In the Wess-Zumino gauge things get simpler, even though one 
should be careful while dealing with classically non gauge-invariant actions. The residual gauge symmetry is 
\begin{equation}
\Xi = \rho -i \theta \bar\theta \partial_+ \rho
\end{equation}
with real $\rho$, while the component expansion of $\mathcal{A}$ and $\mathcal{V}$ read
\begin{subequations}
\begin{align}
\mathcal{A} &= \theta \bar{\theta}^+ A_+ \\
\mathcal{V}  & = A_- -2 i \theta \bar{\mu}_- -2i\bar{\theta}^+ \mu_- + 2 \theta \bar{\theta}^+ D
\end{align}
\end{subequations}
where $D$ is an auxiliary field. Accordingly 
the components $A_{\pm}=A_0 \pm A_1$ of the gauge field are shifted under the residual gauge transformations as
\begin{equation}
A_\pm \stackrel{\rho}{\longrightarrow} A_\pm - \partial_\pm \rho  \\
\end{equation}
The field strength superfield, which is a chiral, is
\begin{equation}
\label{upsidef}
\Upsilon = \bar{D}_+ (\partial_- \mathcal{A}+i \mathcal{V} ) = -2 \left(\mu_- -i \theta (D-i F_{+-} )
-i\theta \bar{\theta}^+ \partial_+ \mu_- \right)
\end{equation}
with $2F_{+-}=\partial_- A_+ - \partial_+ A_-$. We define the gauge-covariant super-derivatives as:
\begin{subequations}
\begin{align}
\mathfrak{D}_+ &= (\partial_{\theta} -i \bar\theta \nabla_+) = D_+ +Q \bar \theta  A_+ \\
\bar{\mathfrak{D}}_+ &= (-\partial_{\bar{\theta}^+} +i \theta \nabla_+) = \bar{D}_+ -Q \theta  A_+ \, .
\end{align}
\end{subequations}
where $\nabla_\pm$ are ordinary covariant derivatives. 

\paragraph{Charged matter multiplets}
A chiral multiplet of charge $Q$ needs to satisfy the gauge-covariant constraint:
\begin{equation}
\bar{\mathfrak{D}}_+ \Phi = 0
\end{equation}
which is solved by
\begin{equation}
\label{chiralcov}
\Phi = \phi + \sqrt{2}\theta \lambda_+ -i \theta \bar\theta  \nabla_+ \phi \, .
\end{equation}
In other words, since 
\begin{equation}
\bar{\mathfrak{D}}_+ = e^{Q \mathcal{A}} \bar{D}_+ e^{-Q \mathcal{A}} 
\end{equation}
We have that 
\begin{equation}
\Phi = e^{Q \mathcal{A}} \Phi_0
\end{equation}
where $\Phi_0$ is a superfield obeying the standard chirality constraint $\bar{D}_+ \Phi_0=0$.

Similarly, a charged Fermi superfield of charge $q$ can be obtained as $\Gamma = e^{q \mathcal{A}} \Gamma_0$ where $\Gamma_0$ 
satisfies $\bar{D}_+ \Gamma_0 = \sqrt{2} E$. Hence the superfield $\Gamma$ has the component expansion:
\begin{equation}
\Gamma =\gamma_- + \sqrt{2} \theta G - \sqrt{2} \bar\theta E(\Phi) -i \theta \bar\theta \nabla_+ \gamma_-\, ,
\end{equation}
where as before $E$ is an holomorphic function in the chiral superfields.


\section{Rational Narain lattices}
\label{app:rat}
 
We discuss in some detail quantization of the two-torus moduli and compatibility of the latter with the two-forms 
$\omega_{1,2}$ characterizing the principal torus bundle. We consider first that there is no Abelian bundle, $i.e.$ that  
'Wilson lines' are turned off. 

Quantization of the torus moduli follows from single-valuelessness of $\exp(iS)$ in any instanton sector~\cite{2011JHEP...03..045A,2013JHEP...11..093I}, or 
from H-flux quantization in supergravity~\cite{Melnikov:2012cv}. It was shown in~\cite{2013JHEP...11..093I} to derive 
from covariance of the model under T-duality along the torus fiber. Moreover 
it was noticed there that these quantization conditions imply that the underlying $c=2$ CFT with 
a two-torus target space is rational.

The Narain Lattice $\Gamma_\textsc{n}\subset \mathbb{R}^{2,2}$ corresponding to the two-torus of metric and B-field:
\begin{equation}
\label{eq:torusmod}
g=\frac{U_{2}}{T_{2}}\begin{pmatrix}1&T_{1}\\T_{1}&|T|^{2}\end{pmatrix}\ , \quad
b=\begin{pmatrix}0&U_{1}\\-U_{1}&0\end{pmatrix}.
\end{equation}
is spanned by 
\begin{align} 
&\frac{1}{\sqrt{2U_2T_2}} \svec{T_2 \\ - T_1 \\ T_2 \\ - T_1} n_1 + 
\frac{1}{\sqrt{2U_2T_2}} \svec{0 \\ 1 \\ 0 \\ 1} n_2 + 
\frac{1}{\sqrt{2U_2T_2}} \svec{U_2 \\ -U_1 \\ -U_2 \\ -U_1} w^1+
\frac{1}{\sqrt{2U_2T_2}} \svec{T_1 U_2 + T_2 U_1 \\ -U_1 T_1 + U_2 T_2 \\ -T_1 U_2+T_2 U_1 \\ -U_1 T_1 - U_2 T_2} w^2 \notag \\
\notag \\
&= e^1 n_1 + e^2 n_2 + \tilde{e}_1 w^1 + \tilde{e}_2 w^2\, .
\label{lattice}
\end{align}
where $n_\ell$ (resp. $w_\ell$) are the integer-valued momenta (resp. winding numbers). The inverse of the 
two-torus metric~(\ref{eq:torusmod}) is $g^{ij} = 2\langle e^i, e^j \rangle\Big|_{\mathbb{R}^{2,0}}$.

The underlying conformal field theory is a rational CFT ($i.e.$ with an extended chiral algebra), if 
$\Gamma_\textsc{l}:=\Gamma_\textsc{n} \cap \mathbb{R}^{2,0}$ and $\Gamma_\textsc{r}:=\Gamma_\textsc{n}\cap \mathbb{R}^{0,2}$ are rank two 
(even and positive-definite) lattices~\cite{Hosono:2002yb}. It is equivalent to require that the space of solutions over 
the integers of the equations  
\begin{subequations}
\label{condrat}
\begin{align}
T_2 n_1 - U_2 w_1 + (U_1 T_2 - T_1 U_2) w_2 &= 0\, ,\\
-T_1 n_1 + n_2 + U_1 w_1 - (U_1 T_1 + U_2 T_2)w_2 &=0 \, .
\end{align}
\end{subequations}
has maximal rank ($i.e.$ rank two). This is satisfied if and only if $U,T \in \mathbb{Q} (\sqrt{D})$, where $D$ is a discriminant of a positive-definite 
even quadratic form, in other words
\begin{equation}
D= b^2 -4 ac <0 \ , \quad a,b,c \in \mathbb{Z} \ , \ a>0 \, .
\end{equation} 

One can eliminate $n_\ell$ in~(\ref{lattice}) using~(\ref{condrat}) and express any element of the lattice $\Gamma_\textsc{l}$ as an element  of the 
'winding lattice' (the sublattice of $\Gamma_\textsc{n}$ defined by $n_1=n_2=0$):
\begin{equation}
\label{eq:windlattice}
p_L = \sfrac{2U_2}{T_2}\mvec{1 \\ 0} w_1 +\sfrac{2U_2}{T_2}\mvec{T_1 \\ T_2} w_2 \, , 
\end{equation}
where, to ensure that it is actually an element of $\Gamma_\textsc{l}$, $w_\ell$ have to satisfy the 'quantization conditions'
\begin{subequations}
\label{eq:quantcondi}
\begin{align}
\frac{U_2}{T_2} (w_1 + T_1 w_2)- U_1 w_2 &\in \mathbb{Z} \,  ,\\
\frac{U_2}{T_2} (T_1 w_1 + |T|^2 w_2)+ U_1 w_1 &\in \mathbb{Z} \, ,
\end{align}
\end{subequations}

The data required to specify a RCFT consists of a 
triple $(\Gamma_{\textsc{l}},\Gamma_{\textsc{r}},\varphi)$, with $\varphi$ being an isometry between the discriminant 
group of the two lattices $\Gamma_{\textsc{l}}$ and $\Gamma_{\textsc{r}}$. The corresponding modular-invariant partition function reads:
\begin{equation}
Z=\frac{1}{\eta^2(\tau) \eta^2 (-\bar \tau)}\sum_{\mu\in\Gamma_{\textsc{l}}^{\star}/\Gamma_{\textsc{l}}}
\Theta_{\mu}^{\Gamma_{\textsc{l}}}\left(\left.\tau\right|0\right)
\Theta_{\varphi(\mu)}^{\Gamma_{\textsc{r}}}\left(\left. -\bar{\tau}\right|0\right)\, .
\end{equation}
where
\begin{equation}
\Theta_{\mu}^{\Gamma}(\tau|\lambda)=
\sum_{\gamma\in\Gamma+\mu}q^{\frac{1}{2}\langle\gamma,\gamma\rangle}e^{2i\pi\langle\gamma,\lambda\rangle} 
\end{equation}
is the theta function with characteristics associated with the lattice $\Gamma$.  $\mu$ is an element of the discriminant 
group $\Gamma^\vee /\Gamma$ of the lattice 
and $\varphi$ is an isometry between the discriminant groups of $\Gamma_\textsc{l}$ and $\Gamma_\textsc{r}$. 

In order to specify the principal two-torus bundle over the $\mathcal{S}$, 
one should further choose two anti-self-dual 
$(1,1)$-forms $\omega_1$ and $\omega_2$ on the $K3$ base $\mathcal{S}$ which define two different integer 
cohomology classes $[\omega_1],[\omega_2]\in H^2(\mathcal{S},\mathbb{Z})$. 
In other words, one should specify a rank-two sublattice $\Gamma_\omega$ of the Picard lattice $\text{Pic}(\mathcal{S})=H^2(\mathcal{S},\mathbb{Z})\cap H^{1,1}_{\bar\partial}(\mathcal{S})$. The metric on this lattice is given by their intersection form: 
\begin{equation}
\label{eq:intform}
\int_\mathcal{S} \omega_i \wedge \omega_j =  d_{ij}\, .
\end{equation}
We remind that the  intersection matrix on the lattice of anti-self dual two-forms on $K3$ can be brought to the form
\begin{equation}
(-E_8)\oplus(-E_8)\oplus -2\begin{pmatrix}1&0&0\\0&1&0\\0&0&1\end{pmatrix}\, .
\end{equation}

One is thus endowed with two quadratic even lattices $\Gamma_{\textsc{l}}$ and $\Gamma_\omega$. We define first the following element of a formal extension of the winding lattice, 
valued in $H^2 (\mathcal{S},\mathbb{Z})\times H^2 (\mathcal{S},\mathbb{Z})$, as:
\begin{equation}
p_\omega = \sfrac{2U_2}{T_2}\mvec{1 \\ 0} \omega_1 +\sfrac{2U_2}{T_2}\mvec{T_1 \\ T_2} \omega_2 \, .
\end{equation}
One should impose  that $p_\omega$ belongs actually to a formal extension $\Gamma_\textsc{l}\otimes\text{Pic}(\mathcal{S})$ of the left lattice, $i.e.$ one should 
impose 'compatibility conditions', of the same form as~(\ref{eq:quantcondi}):
\begin{subequations}
\label{eq:quantcondforms}
\begin{align}
\frac{U_2}{T_2} (\omega_1 + T_1 \omega_2)- U_1 \omega_2 &\in H^2 (\mathcal{S},\mathbb{Z}) \,  ,\\
\frac{U_2}{T_2} (T_1 \omega_1 + |T|^2 \omega_2)+ U_1 \omega_1 &\in H^2 (\mathcal{S},\mathbb{Z}) \,  .
\end{align}
\end{subequations}
Importantly, these conditions depends on $\Gamma_\textsc{n}$ and not of $\Gamma_\textsc{l}$ only.

To summarize, for a given pair of anti-self dual two-forms and $(\omega_1,\omega_2)$, defining a rank two (generically) lattice 
$\Gamma_\omega \in H^2 (\mathcal{S},\mathbb{Z})$, one needs to choose the metric and B-field of the two torus such that two conditions are satisfied:
\begin{enumerate}
\item The compatibility condition~(\ref{eq:quantcondforms}),
\item The tadpole condition $N-24 = \int_\mathcal{S} \langle p_\omega,p_\omega\rangle$, where $N=-\int_\mathcal{S} \text{ch}_2 (\mathcal{V})$ is the instanton number.
\end{enumerate}

In order to include Abelian bundles, or 'Wilson lines', one first consider an embedding of the toroidal $(2,2)$ lattice into a $(10,2)$ lattice which includes also the contribution from the $E_8$ weight 
lattice.  The moduli $V^a$ in eq.~(\ref{eq:connection}) corresponds then to off-diagonal deformations of this lattice. Quantization of $(T,U)$, as well as the compatibility conditions~(\ref{eq:quantcondforms}), are not affected. The 'physical' two-torus metric 
is not given anymore by eq.~(\ref{eq:torusmod}), but is of the form  $g_{IJ} + \frac{\alpha'}{4} \text{Tr} (A_I A_J)$, 
as for ordinary Wilson lines, however the tadpole and compatibility conditions remains unchanged.

\subsubsection*{An example}

For illustration let us consider an orthogonal torus with radii $R_1 = \sqrt{p_1 /q_1}$ and 
$R_2 = \sqrt{p_2 /q_2}$ hence 
\begin{equation}
U =i\sfrac{p_1 p_2}{q_1 q_2} \ , \quad T = i \sfrac{p_2 q_1}{q_2 p_1} \, .
\label{eq:ortholat}
\end{equation}
The quantization conditions~(\ref{eq:quantcondi}) read then
\begin{equation}
p_1 w_1 \equiv 0 \mod q_1 \ , \quad p_2 w_2 \equiv 0 \mod q_2 \, .
\end{equation}
Assuming that $\text{gcd}\, (q_\ell,p_\ell)=1$ this is solved by choosing 
$w_1 = q_1 W_1$ and $w_2 = q_2 W_2$. Elements of the lattice $\Gamma_\textsc{l}$ are then of the form 
\begin{equation}
\label{eq:ortholattice}
p_L = \sqrt{2 p_1 q_1} \mvec{1 \\ 0} W_1 +\sqrt{2p_2 q_2}\mvec{0 \\ 1} W_2 \, ,
\end{equation}
with $W_\ell \in \mathbb{Z}$. A basis of $\Gamma_\textsc{l}$ is then provided by 
\begin{equation}
e^\textsc{l}_1 = \sqrt{2 p_1 q_1} \mvec{1 \\ 0} \ , \quad 
e^\textsc{l}_2 = \sqrt{2p_2 q_2}\mvec{0 \\ 1}\, .
\end{equation}

The modular-invariant partition function associated with this rational Narain lattice reads:
\begin{equation}
Z = 
\frac{1}{\eta^2(\tau) \eta^2 (-\bar \tau)} \prod_{\ell=1}^2 
\sum_{r_\ell \in \mathbb{Z}_{2p_\ell}} \sum_{s_\ell \in \mathbb{Z}_{2q_\ell}} 
\Theta_{q_\ell r_\ell + p_\ell s_\ell, 2 p_\ell q_\ell}\left(\left.\tau\right|0\right)
\Theta_{q_\ell r_\ell - p_\ell s_\ell, 2 p_\ell q_\ell}\left(\left. -\bar{\tau}\right|0\right)\, .
\label{eq:pftor}
\end{equation}
The isometry $\varphi~: \Gamma_\textsc{L}^\star/ \Gamma_\textsc{L} \to \Gamma_\textsc{r}^\star/ \Gamma_\textsc{r}$ can 
be determined explicitly by  mapping $q_\ell r_\ell \pm p_\ell s_\ell$ into the 'fundamental domain' 
$\{0,\ldots,2p_\ell q_\ell -1\}$. 

The compatibility condition between $\Gamma_\omega$ and the orthogonal lattice of 
moduli~(\ref{eq:ortholat}) amounts to 
\begin{equation}
\omega_\textsc{l}^\ell := \frac{1}{q_\ell}\,  \omega^\ell \in H^2 (\mathcal{S},\mathbb{Z})\, ,\ \ell =1,2 \, .
\end{equation}

Notice that the partition function~(\ref{eq:pftor}) differs from the partition function for an 
orthogonal torus with radii $R_\ell = \sqrt{2 p_\ell q_\ell}$ precisely in the 
choice of the isometry $\varphi$. In the latter case one has indeed simply
\begin{equation}
Z = \frac{1}{\eta^2(\tau) \eta^2 (-\bar \tau)} \prod_{\ell=1}^2 
\sum_{m_\ell \in \mathbb{Z}_{2p_\ell q_\ell}} 
\Theta_{m_\ell, 2 p_\ell q_\ell}\left(\left.\tau\right|0\right)
\Theta_{m_\ell, 2 p_\ell q_\ell}\left(\left. -\bar{\tau}\right|0\right)\, .
\end{equation}
Moreover in this case the chiral lattices $\Gamma_\textsc{l}$ and 
$\Gamma_\textsc{r}$ coincide such that the compatiblity condition is trivial. 

One can also consider examples obtained from the above by T-duality along the two circles, each of these 
cases corresponding to a different choice of isometry $\varphi$. Satisfying 
the compatibility condition \cref{eq:ortholat} is then equivalent to considering in each case a different sublattice of $\text{Pic}(\mathcal{S})$. This 
is one of the dualities studied in~\cite{2013JHEP...11..093I}, which induces a duality action on $\text{Pic}(\mathcal{S})$, leaving the tadpole condition invariant 
by construction. 

\section{Proof of the geometrical formula}
\label{app:ProofGeom}

In this appendix we give a proof of the geometrical formula \cref{eq:geom} in the case where the $K3$ base $\mathcal{S}$ is constructed as a subvariety of a projective space $\mathbb{P}^n$ or more 
generally of a weighted projective space $V=\mathbb{P}^n(q_0,...,q_n)$. We restrict to the case where the subvariety do not intersect the singular loci of the ambient space.

One has the following dual of the normal bundle sequence:
\begin{equation}
0\rightarrow N_{\mathcal{S}/V}^\star\rightarrow T^\star_V|_\mathcal{S}\rightarrow T^\star_\mathcal{S}\rightarrow 0\, ,
\end{equation}
which gives the following long exact sequence in sheaf cohomology:
\begin{align}
\dots&\rightarrow H^1(\mathcal{S},N_{\mathcal{S}/V}^\star)\overset{\alpha_1}{\rightarrow} H^1(\mathcal{S},T^\star_V|_\mathcal{S})\rightarrow H^1(\mathcal{S},T^\star_\mathcal{S})\rightarrow\nonumber\\
 &\rightarrow H^2(\mathcal{S},N_{\mathcal{S}/V}^\star)\overset{\alpha_2}{\rightarrow} H^2(\mathcal{S},T^\star_V|_\mathcal{S})\rightarrow H^2(\mathcal{S},T^\star_\mathcal{S})\rightarrow\dots \, .
\end{align}
The exactness of this sequence gives in particular:
\begin{equation}
H^{1,1}_{\bar\partial}(\mathcal{S})\simeq \text{coker}(\alpha_1)\oplus\text{ker}(\alpha_2)\, .
\end{equation}
For the proof we will restrict to the case where the hypersurface $\mathcal{S}$ is \textit{favourable}, namely that $\text{ker}(\alpha_2)$ is trivial and $\alpha_1$ is surjective. In this case, all 
the elements of $H^{1,1}_{\bar\partial}(\mathcal{S})$ can be understood as being inherited from the ambient space. Moreover, since for a weighted projective space one has 
$\text{Pic}(\mathbb{P}^n(q_0,...,q_n))=\mathbb{Z}$ as a finitely generated abelian group, we have that $\text{rk}(\text{Pic}(\mathcal{S}))=1$\footnote{This actually implies that one restricts to examples 
which are non-supersymmetric in spacetime, since the two-form $\omega$ then fails to be primitive with respect to the base. However, this restriction can be straightforwardly overcome, see the end of this appendix. Moreover, neither 
from the two-dimensional QFT  nor from the mathematical viewpoint this seems to play an important role.}.
Following the mathematical literature and given two holomorphic vector bundles $E$ and $F$ over $\mathcal{S}$, we define the formal series with bundle coefficients:
\begin{equation}
\mathbb{E}_{q,y}(E,F):=\bigotimes_{n=1}^{\infty}\left(\bigwedge\nolimits_{-y q^{n-1}}F^\star\otimes\bigwedge\nolimits_{-y^{-1} q^{n}}F\otimes S_{q^{n}}E^\star\otimes S_{q^{n}}E\right)\, .
\end{equation}
We also consider a heterotic Narain lattice $\Gamma(T,U,V)$ with a left coupling to $p_\omega$  characterizing 
the torus bundle, see eq.~(\ref{eq:p_omega_def}), of partition function:
\begin{equation}
\mathcal{Z}(\tau,\bar\tau,p_\omega)=\sum_{(p_\textsc{l},p_\textsc{r})
\in\Gamma_{10,2}}\frac{q^{\frac{1}{4}|p_\textsc{l}|^2}}{\eta(\tau)^{18}}\frac{\bar q^{\frac{1}{4}|p_\textsc{r}|^2}}{\bar\eta(\bar\tau)^2}
\exp\left(-2i\pi \, \text{Re}\left(p_\omega\, \overline{p_\textsc{l}^0}\right)\right)\, ,
\end{equation}
Given this data, and for $\mathcal{V}$ a holomorphic vector bundle of rank $r$ over $\mathcal{S}$, we introduce the following modified holomorphic Euler characteristic:
\begin{equation}
\chi(\mathcal{S},\mathcal{V},\omega):=q^{\frac{r-(n-1)}{12}}y^{-\frac{r}{2}}\int_\mathcal{S}\text{ch}(\mathbb{E}_{q,y}(T_\mathcal{S},\mathcal{V}))\, \text{td}(T_\mathcal{S})\, \mathcal{Z}(\tau,\bar\tau,\omega)\, .
\end{equation}
Given a holomorphic vector bundles $E$ of formal Chern roots $\{x_a\}$, let us define the following objects:
\begin{subequations}
\begin{align}
f(E)&:=\text{ch}\left(\bigotimes_{n=1}^{\infty}\left(\bigwedge\nolimits_{-y q^{n-1}}E^\star\otimes\bigwedge\nolimits_{-y^{-1} q^{n}}E\right)\right)\, ,\\
g(E)&:=\text{ch}\left(\bigotimes_{n=1}^{\infty}\left(S_{q^{n}}E^\star\otimes S_{q^{n}}E\right)\right)\text{td}\left(E\right)\, .
\end{align}
\end{subequations}
Using the total Chern characters for the total symmetric and skew-symmetric products:
\begin{equation}
\text{ch}(S_t E)=\prod_a\frac{1}{1-te^{x_a}}\, ,\ \ \ \ \text{ch}\left(\bigwedge\nolimits_{t} E\right)=\prod_a\left(1+te^{x_a}\right)\, ,
\end{equation}
one can show that:
\begin{subequations}
\begin{align}
f(E)&=\prod_a \left\{\prod_{n=1}^{\infty}\left(1-yq^{n-1}e^{-x_a}\right)\left(1-y^{-1}q^{n}e^{x_a}\right)\right\}\, ,\\
g(E)&=\prod_a \left\{x_a\prod_{n=1}^{\infty}\frac{1}{\left(1-q^{n-1}e^{-x_a}\right)\left(1-q^{n}e^{x_a}\right)}\right\}\, ,
\end{align}
\end{subequations}
leading to:
\begin{subequations}
\begin{align}
f(E)&=\prod_a \left\{q^{-1/12}y^{1/2}e^{-x_a/2}\, \frac{i\theta_1\left(\tau\left|\frac{x_a}{2i\pi}-z\right.\right)}{\eta(\tau)}\right\}\, ,\\
g(E)&=\prod_a \left\{q^{1/12}e^{x_a/2}\, \frac{\eta(\tau)\, x_a}{i\theta_1\left(\tau\left|\frac{x_a}{2i\pi}\right.\right)}\right\}\, .
\end{align}
\end{subequations}
In our context one has the following defining short exact sequences for the hypersurface $\mathcal{S}$ and the rank $r$ holomorphic vector bundle $\mathcal{V}$ over it:
\begin{subequations}
\begin{align}
&0\rightarrow T_\mathcal{S}\rightarrow T_V|_\mathcal{S}\rightarrow \mathcal{O}_V(k)|_\mathcal{S}\rightarrow 0\, ,\\
&0\rightarrow \mathcal{V}\rightarrow \bigoplus_{a=0}^{r}\mathcal{O}(Q_a)|_\mathcal{S}\overset{\otimes J^a}{\rightarrow}\mathcal{O}(-Q_P)|_\mathcal{S}\rightarrow 0\, .
\end{align}
\end{subequations}
Using multiplicative properties of $f$ and $g$, one obtains:
\begin{equation}
f\left(\bigoplus_{a=0}^{r}\mathcal{O}(Q_a)\right)=f(\mathcal{V})f(\mathcal{O}(-Q_P))\, ,\ \ \ \ \ g(T_V|_\mathcal{S})=g(T_\mathcal{S})g(\mathcal{O}_V(k)|_\mathcal{S})\, .
\end{equation}
The formal Chern roots are defined through the following total Chern classes:
\begin{equation}
c(\mathcal{O}(m))=1+mH\, ,\ \ \ \ \ c(V)=\prod_{i=0}^n (1+q_iH)\, ,
\end{equation}
leading to:
\begin{equation}
f(\mathcal{V})\left.\left[\frac{i\theta_1\left(\tau\left|-\frac{Q_PH}{2i\pi}-z\right.\right)}{\eta(\tau)}\right]\right|_\mathcal{S}=q^{-r/12}y^{r/2}
e^{-\left(\sum Q_a+Q_P\right)\frac{H}{2}}\prod_{a=0}^{r}\left.\left[\frac{i\theta_1\left(\tau\left|\frac{Q_aH}{2i\pi}-z\right.\right)}{\eta(\tau)}\right]\right|_\mathcal{S}\, ,
\end{equation}
Exploiting the Euler exact sequence:
\begin{equation}
0\rightarrow \mathcal{O}_V\rightarrow \bigoplus_{i=0}^n\mathcal{O}_V(q_i)\rightarrow T_V\rightarrow 0\, ,
\end{equation}
one also has:
\begin{equation}
g(T_\mathcal{S})\left.\left[\frac{\eta(\tau)\, kH}{i\theta_1\left(\tau\left|\frac{kH}{2i\pi}\right.\right)}\right]\right|_\mathcal{S}=q^{\frac{n-1}{12}}
\eta(\tau)^2\prod_{i=0}^{n}\left.\left[\frac{\eta(\tau)\, q_iH}{i\theta_1\left(\tau\left|\frac{q_iH}{2i\pi}\right.\right)}\right]\right|_\mathcal{S}\, .
\end{equation}
The two equations above give the 
contributions from the tangent bundle and from the holomorphic vector bundle $\mathcal{V}$ in terms of the embedding in the ambient space $V$. Concerning the lattice part, $\text{Pic}(\mathcal{S})$ 
being of rank one, one necessarily has $\omega_\ell=m_\ell H$. 

Turning the integral over the hypersurface to an integral over the ambient space via:
\begin{equation}
\int_\mathcal{S}\varphi=\int_V c_1(\mathcal{O}_V(k))\varphi\, ,
\end{equation}
One obtains:
\begin{align}
\chi(\mathcal{S},\mathcal{V},\omega)=&\, 
\eta(\tau)^2\prod_{i=0}^{n}q_i\int_V\, H^{n+1}\, \prod_{a=0}^{r}\left[\frac{i\theta_1\left(\tau\left|\frac{Q_aH}{2i\pi}-z\right.\right)}{\eta(\tau)}\right]
\left[\frac{\eta(\tau)}{i\theta_1\left(\tau\left|-\frac{Q_PH}{2i\pi}-z\right.\right)}\right]\nonumber\\
&\prod_{i=0}^{n}\left[\frac{\eta(\tau)}{i\theta_1\left(\tau\left|\frac{q_iH}{2i\pi}\right.\right)}\right]
\left[\frac{i\theta_1\left(\tau\left|\frac{kH}{2i\pi}\right.\right)}{\eta(\tau)}\right]\, \mathcal{Z}(\tau,\bar\tau,\mathfrak{m}H)\, .
\end{align}
Using the fact that the hyperplane class is normalized such that:
\begin{equation}
\prod_i q_i\int_V H^n=1\, ,
\end{equation}
together with the residue formula, to turn $\int_V H^{n+1}\varphi(H)$ into a contour integral around the origin of the complex plane 
$\oint_{u=0}\text{d}u\, \varphi(2i\pi u)$, one concludes that this modified holomorphic Euler characteristic coincides with 
the formula that would be obtained in the geometrical phase of the torsional GLSM, $i.e.$ considering the contour integral around $u=0$ in 
eq.~(\ref{eq:oneloopdetrank}).

This proof can be generalized by allowing for non-favorable hypersurfaces, or by considering complete intersections in more generic toric varieties as ambient spaces. One would then 
end up with higher dimensional residue operations, leading to Jeffrey-Kirwan residue formul\ae{} type. The generalization to the non-toric case is left for later work.


\bibliographystyle{JHEP}
\bibliography{biblioJ}

\end{document}